\documentclass[english,american]{IEEEtran}
\usepackage[T1]{fontenc}
\usepackage[latin9]{inputenc}
\usepackage[letterpaper]{geometry}
\usepackage{color}
\usepackage{verbatim}
\usepackage{float}
\usepackage{calc}
\usepackage{amsthm}
\usepackage{amsmath}
\usepackage{graphicx}
\usepackage{amssymb}

\makeatletter

\floatstyle{ruled}
\newfloat{algorithm}{tbp}{loa}
\floatname{algorithm}{Algorithm}

\theoremstyle{plain}
\newtheorem{thm}{Theorem}
\theoremstyle{definition}
\newtheorem{defn}[thm]{Definition}
\theoremstyle{plain}
\newtheorem{cor}[thm]{Corollary}
\theoremstyle{remark}
\newtheorem{rem}[thm]{Remark}
\theoremstyle{definition}
\newtheorem{example}[thm]{Example}
\theoremstyle{plain}
\newtheorem{lem}[thm]{Lemma}
\theoremstyle{plain}
\newtheorem{prop}[thm]{Proposition}

 \usepackage{bbm}

\usepackage{verbatim}
\usepackage{alltt}
\usepackage{float}
 \usepackage[hang,small,bf]{caption}

\DeclareMathOperator*{\argmin}{arg\,min}
\DeclareMathOperator*{\argmax}{arg\,max}
\date{}

\floatstyle{ruled}
\restylefloat{table}

\floatname{table}{Problem}

\IEEEoverridecommandlockouts

\makeatother

\usepackage{babel}

\begin{document}

\title{Joint Decoding of LDPC Codes and Finite-State Channels via Linear-Programming}

\author{Byung-Hak Kim{*}, \emph{Student Member, IEEE} and Henry D. Pfister,
\emph{Senior Member, IEEE}%
\thanks{This material is based upon work supported by the National Science
Foundation under Grant No. 0747470. The material in this paper was
presented in part at IEEE International Symposium on Information Theory
(ISIT), Austin, TX, June 2010 and the IEEE International Conference
on Communications (ICC), Kyoto, Japan, June 2011. 

The authors are with the Department of Electrical and Computer Engineering,
Texas A\&M University, College Station, TX 77843, USA (email: bhkim@tamu.edu;
hpfister@tamu.edu).%
}}
\maketitle
\begin{abstract}
This paper considers the joint-%
\begin{comment}
iterative 
\end{comment}
{}decoding problem for finite-state channels (FSCs) and low-density
parity-check (LDPC) codes. In the first part, the linear-programming
(LP) decoder for binary linear codes%
\begin{comment}
, introduced by Feldman et al.
\end{comment}
{} is extended to joint-decoding of binary-input FSCs. In particular,
we provide a rigorous definition of LP joint-decoding pseudo-codewords
(JD-PCWs) that enables evaluation of the pairwise error probability
between codewords and JD-PCWs in AWGN. This leads naturally to a provable
upper bound on decoder failure probability. If the channel is a finite-state
intersymbol interference channel, then the joint LP decoder also has
the maximum-likelihood (ML) certificate property and all integer-valued
solutions are codewords. In this case, the performance loss relative
to ML decoding can be explained completely by fractional-valued JD-PCWs.
After deriving these results, we discovered some elements were equivalent
to earlier work by Flanagan on linear-programming receivers.

In the second part, we develop an efficient iterative solver for the
joint LP decoder discussed in the first part. In particular, we extend
the approach of iterative approximate LP decoding, proposed by Vontobel
and Koetter and analyzed by Burshtein, to this problem. By taking
advantage of the dual-domain structure of the joint-decoding LP, we
obtain a convergent iterative algorithm for joint LP decoding whose
structure is similar to BCJR-based turbo equalization (TE). The result
is a joint iterative decoder whose per-iteration complexity is similar
to that of TE but whose performance is similar to that of joint LP
decoding. The main advantage of this decoder is that it appears to
provide the predictability of joint LP decoding and superior performance
with the computational complexity of TE. One expected application
is coding for magnetic storage where the required block-error rate
is extremely low and system performance is difficult to verify by
simulation.\end{abstract}

\section{Introduction \label{sec:Intro} }

\subsection{Motivation and Problem Statement}

Iterative decoding of error-correcting codes, while introduced by
Gallager in his 1960 Ph.D. thesis, was largely forgotten until the
1993 discovery of turbo codes by Berrou et al. Since then, message-passing
iterative decoding has been a very popular decoding algorithm in research
and practice. In 1995, the turbo decoding of a finite-state channel
(FSC) and a convolutional code (instead of two convolutional codes)
was introduced by Douillard et al. as \emph{turbo equalization} (TE)
and this enabled the joint-decoding of the channel and the code by
iterating between these two decoders \cite{Douillard-ett95}. Before
this, one typically separated channel decoding (i.e., estimating the
channel inputs from the channel outputs) from the decoding of the
error-correcting code (i.e., estimating the transmitted codeword from
estimates of the channel inputs) \cite{Gallager-60}\cite{Muller-it04}.
This breakthrough received immediate interest from the magnetic recording
community, and TE was applied to magnetic recording channels by a
variety of authors (e.g., \cite{Ryan-icc98,McPheters-asilo98,Oberg-aller98,Tuchler-com02}).
TE was later combined with turbo codes and also extended to low-density
parity-check (LDPC) codes (and called \emph{joint iterative decoding})
by constructing one large graph representing the constraints of both
the channel and the code (e.g., \cite{Kurkoski-it02,FCR-2004}).

In the magnetic storage industry, error correction based on Reed-Solomon
codes with hard-decision decoding has prevailed for the last 25 years.
Recently, LDPC codes have attracted a lot of attention and some hard-disk
drives (HDDs) have started using iterative decoding (e.g., \cite{Dholakia-commag04,Anastasopoulos-07,Kavcic-08}).
Despite progress in the area of reduced-complexity detection and decoding
algorithms, there has been some resistance to the deployment of TE
structures (with iterative detectors/decoders) in magnetic recording
systems because of error floors and the difficulty of accurately predicting
performance at very low error rates. Furthermore, some of the spectacular
gains of iterative coding schemes have been observed only in simulations
with block-error rates above $10^{-6}.$ The challenge of predicting
the onset of error floors and the performance at very low error rates,
such as those that constitute the operating point of HDDs (the current
requirement of an overall block error rate of $10^{-12})$, remains
an open problem. The presence of error floors and the lack of analytical
tools to predict performance at very low error rates are current impediments
to the application of iterative coding schemes in magnetic recording
systems.

In the last five years, linear programming (LP) decoding has been
a popular topic in coding theory and has given new insight into the
analysis of iterative decoding algorithms and their modes of failure
\cite{Feldman-2003}\cite{Feldman-it05}\cite{Pseudocodewords-web}.
In particular, it has been observed that LP decoding sometimes performs
better than iterative (e.g., sum-product) decoding in the error-floor
region. We believe this stems from the fact that the LP decoder always
converges to a well-defined LP optimum point and either detects decoding
failure or outputs an ML codeword. For both decoders, fractional vectors,
known as pseudo-codewords (PCWs), play an important role in the performance
characterization of these decoders \cite{Feldman-it05}\cite{Vontobel-arxiv05}.
This is in contrast to classical coding theory where the performance
of most decoding algorithms (e.g., maximum-likelihood (ML) decoding)
is completely characterized by the set of codewords. 

While TE-based joint iterative decoding provides good performance
close to capacity, it typically has some trouble reaching the low
error rates required by magnetic recording and optical communication.
To combat this, we extend the LP decoding to the joint-decoding of
a binary-input FSC and an outer LDPC code. During the review process
of our conference paper on this topic \cite{Kim-isit10}, we discovered
that this LP formulation is mathematically equivalent to Flanagan's
general formulation of linear-programming receivers \cite{Flanagan-aller08,Flanagan-arxiv09}.
Since our main focus was different than Flanagan's, our main results
and extensions differ somewhat from his. In particular, our main motivation
is that critical storage applications (e.g., HDDs) require block error
rates that are too low to be easily verifiable by simulation. For
these applications, \textcolor{black}{an efficient iterative solver
for the joint-decoding LP} would have favorable properties: error
floors predictable by pseudo-codeword analysis and convergence based
on a well-defined optimization problem. Therefore, we introduce a
novel iterative solver for the joint LP decoding problem whose per-iteration
complexity (e.g., memory and time) is similar to that of TE but whose
performance appears to be superior at high SNR \cite{Kim-isit10}\cite{Kim-icc11}.

\subsection{Notation}

Throughout the paper we borrow notation from \cite{Feldman-it05}.
Let $\mathcal{I}=\left\{ 1,\,\ldots,\, N\right\} $ and $\mathcal{J}=\left\{ 1,\,\ldots,\, M\right\} $
be sets of indices for the variable and parity-check nodes of a binary
linear code. A variable node $i\in\mathcal{I}$ is connected to the
set $\mathcal{N}(i)$ of neighboring parity-check nodes. Abusing notation,
we also let $\mathcal{N}(j)$ be the neighboring variable nodes of
a parity-check node $j\in\mathcal{J}$ when it is clear from the context.
For the trellis associated with a FSC, we let $E=\left\{ 1,\,\ldots,\, O\right\} $
index the set of trellis edges associated with one trellis section,
$\mathcal{S}$ be the set of possible states, and $\mathcal{A}$ be
the possible set of noiseless output symbols. For each edge%
\footnote{In this paper, $e$ is used to denote a trellis edge while $\texttt{e}$
denotes the universal constant that satisfies $\ln\texttt{e}=1$.%
}, $e\in E^{N}$, in the length-$N$ trellis, the functions $t:E^{N}\rightarrow\{1,\ldots,N\}$,~$s:E^{N}\rightarrow\mathcal{S}$,
$s':E^{N}\rightarrow\mathcal{S}$,~$x:E^{N}\rightarrow\{0,1\}$,~and
$a:E^{N}\rightarrow\mathcal{A}$ map this edge to its respective time
index, initial state, final state, input bit, and noiseless output
symbol. Finally, the set of edges in the trellis section associated
with time $i$ is defined to be $\mathcal{T}_{i}=\left\{ e\in E^{N}\,|\, t(e)=i\right\} $.

\subsection{Background: LP Decoding and Finite-State Channels \label{sec:LPD}}

In \cite{Feldman-2003}\cite{Feldman-it05}, Feldman et al. introduced
a linear-programming (LP) decoder for binary linear codes, and applied
it specifically to both LDPC and turbo codes. It is based on solving
an LP relaxation of an integer program that is equivalent to maximum-likelihood
(ML) decoding. For long codes and/or low SNR, the performance of LP
decoding appears to be slightly inferior to belief-propagation decoding.
Unlike the iterative decoder, however, the LP decoder either detects
a failure or outputs a codeword which is guaranteed to be the ML codeword. 

Let $\mathcal{C}\subseteq\left\{ 0,1\right\} ^{N}$ be the length-$N$
binary linear code defined by a parity-check matrix and $\mathbf{c}=(c_{1},\ldots,c_{N})$
be a codeword. Let $\mathcal{L}$ be the set whose elements are the
sets of indices involved in each parity check, or \[
\mathcal{L}=\left\{ \mathcal{N}(j)\subseteq\{1,\ldots,N\}|\, j\in\mathcal{J}\right\} .\]
Then, we can define the set of codewords to be\[
\mathcal{C}=\left\{ \mathbf{c}\in\left\{ 0,1\right\} ^{N}\,\bigg|\,\sum_{i\in L}c_{i}\equiv0\,\bmod2,\,\forall\, L\in\mathcal{L}\right\} .\]
The \emph{codeword polytope} is the convex hull of $\mathcal{C}$.
This polytope can be quite complicated to describe though, so instead
one constructs a simpler polytope using local constraints. Each parity-check
$L\in\mathcal{L}$ defines a local constraint equivalent to the extreme
points of a polytope in $\left[0,1\right]^{N}$.
\begin{defn}
\label{def:LCP} The \emph{local codeword polytope} $\mbox{LCP(\ensuremath{L})}$
associated with a parity check is the convex hull of the bit sequences
that satisfy the check. It is given explicitly by\[
\mbox{LCP}(L)\triangleq\!\!\bigcap_{\substack{S\subseteq L\\
\left|S\right|\text{odd}}
}\left\{ \mathbf{c}\in[0,\,1]^{N}\,\bigg|\sum_{i\in S}c_{i}-\!\!\!\sum_{i\in L-S}\!\! c_{i}\leq\left|S\right|\!-\!1\right\} .\]

\end{defn}
We use the notation $\mathcal{P}(H)$ to denote the simpler polytope
corresponding to the intersection of local check constraints; the
formal definition follows.
\begin{defn}
The \emph{relaxed polytope} $\mathcal{P}(H)$ is the intersection
of the LCPs over all checks and \begin{align*}
\mathcal{P}(H) & \triangleq\bigcap_{L\in\mathcal{L}}\mbox{LCP}(L).\end{align*}

\end{defn}
The LP decoder and its ML certificate property is characterized by
the following theorem.
\begin{thm}
[\cite{Feldman-2003}] \emph{Consider $N$ consecutive uses of a symmetric
channel $\Pr\left(Y=y|C=c\right)$. If a uniform random codeword is
transmitted and $\mathbf{y}=(y_{1},\ldots,y_{N})$ is received, then
the LP decoder outputs $\mathbf{f}=(f_{1},\ldots,f_{N})$ given by
\[
\argmin_{\mathbf{f}\in\mathcal{P}(H)}\sum_{i=1}^{N}f_{i}\,\ln\left(\frac{\mbox{Pr}(Y_{i}=y_{i}\,|\, C_{i}=0)}{\mbox{Pr}(Y_{i}=y_{i}\,|\, C_{i}=1)}\right),\]
which is the ML solution if $\mathbf{f}$ is integral (i.e., $\mathbf{f}\in\left\{ 0,1\right\} ^{N}$).} 
\end{thm}
From simple LP-based arguments, one can see that LP decoder may also
output nonintegral solutions.
\begin{defn}
An \emph{LP decoding pseudo-codeword} (LPD-PCW) of a code defined
by the parity-check matrix $H$ is any \emph{nonintegral} vertex of
the relaxed (fundamental) polytope $\mathcal{P}(H).$ 
\end{defn}
We also define the finite-state channel, which can be seen as a model
for communication systems with memory where each output depends only
on the current input and the previous channel state instead of the
entire past.
\begin{defn}
A \emph{finite-state channel} (FSC) defines a probabilistic mapping
from a sequence of inputs to a sequence of outputs. Each output $Y_{i}\in\mathcal{Y}$
depends only on the current input $X_{i}\in\mathcal{X}$ and the previous
channel state $S_{i-1}\in\mathcal{S}$ instead of the entire history
of inputs and channel states. Mathematically, we define $P\left(y,s'|x,s\right)\triangleq\mbox{Pr}\left(Y_{i}\!=\! y,S_{i}\!=\! s'|X_{i}\!=\! x,S_{i-1}\!=\! s\right)$
for all $i$, and use the shorthand notation $P_{0}(s)\triangleq\Pr(S_{0}=s)$
and \begin{align*}
\! P\!\left(y_{1}^{N}\!,s_{1}^{N}|x_{1}^{N}\!,s_{0}\right) & \!\triangleq\!\mbox{Pr}\left(Y_{1}^{N}\!\!=\! y_{1}^{N}\!,S_{1}^{N}\!\!=\! s_{1}^{N}|X_{1}^{N}\!\!=\! x_{1}^{N}\!,S_{0}\!=\! s_{0}\right)\\
 & \!=\!\prod_{i=1}^{N}P\left(y_{i},s_{i}|x_{i},s_{i-1}\right),\end{align*}
where the notation $Y_{i}^{j}$ denotes the subvector $(Y_{i},Y_{i+1},\ldots,Y_{j})$.
\end{defn}
An important subclass of FSCs is the set of finite-state intersymbol
interference channels which includes all deterministic finite-state
mappings of the inputs corrupted by memoryless noise.
\begin{defn}
\label{def:FSISI}A \emph{finite-state intersymbol interference channel}
(FSISIC) is a FSC whose next state is a deterministic function, $\eta(x,s)$,
of the current state $s$ and input $x$. Mathematically, this implies
that\[
\sum_{y\in\mathcal{Y}}P\left(y,s'|x,s\right)=\begin{cases}
1 & \mbox{if}\,\eta(x,s)=s'\\
0 & \mbox{otherwise}\end{cases}.\]

\end{defn}
Though our derivations are general, we use the following FSISIC examples
throughout the paper to illustrate concepts and perform simulations.
\begin{figure}[t]
\begin{centering}
\includegraphics[scale=0.65]{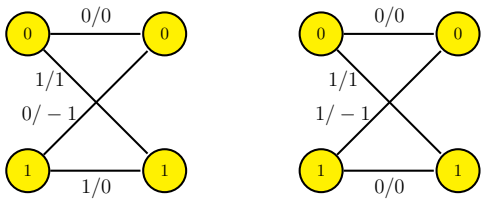}
\par\end{centering}

\caption{\label{fig:dic}State diagrams for noiseless dicode channel without
(left) and with precoding (right). The edges are labeled by the input/output
pair.}

\end{figure}

\begin{defn}
\label{def:exFSISI}The \emph{dicode channel} (DIC) is a binary-input
FSISIC with an impulse response of $G(z)=1-z^{-1}$ and additive Gaussian
noise \cite{Immink-it98}. If the input bits are differentially encoded
prior to transmission, then the resulting channel is called the \emph{precoded
dicode channel} (pDIC) \cite{Immink-it98}. The state diagrams of
these two channels are shown in Fig. \ref{fig:dic}. For the trellis
associated with a DIC and pDIC, we let $E=\left\{ 1,2,3,4\right\} ,\,\mathcal{S}=\left\{ 0,1\right\} $
and $\mathcal{A}=\left\{ -1,0,1\right\} .$ Also, the \emph{class-II
Partial Response} (PR2) channel is a binary-input FSISIC with an impulse
response of $G(z)=1+2z^{-1}+z^{-2}$ and additive Gaussian noise \cite{Immink-it98}\cite{Jeon-itm07}.
\end{defn}

\subsection{Outline of the Paper}

The remainder of the paper is organized as follows. In Section \ref{sec:Joint-LPD},
we introduce the joint LP decoder, define joint-decoding pseudo-codewords
(JD-PCWs), and describe the appropriate generalized Euclidean distance
for this problem. Then, we discuss the decoder performance analysis
using the union bound (via pairwise error probability) over JD-PCWs.
Section \ref{sec:IterativeJoint-LPD} is devoted to developing the
iterative solver for the joint LP decoder, i.e., iterative joint LP
decoder and its proof of convergence. %
\begin{comment}
 and further extend them to universal iterative joint LP decoder for
unknown FSISI case.
\end{comment}
{} Finally, Section \ref{sec:Simulations} presents the decoder simulation
results and Section \ref{sec:Concl} gives some conclusions.

\section{Joint LP Decoder \label{sec:Joint-LPD}}

Feldman et al. introduced the LP decoder for binary linear codes in
\cite{Feldman-2003}\cite{Feldman-it05}. It is is based on an LP
relaxation of an integer program that is equivalent to ML decoding.
Later, this method was extended to codes over larger alphabets \cite{Flanagan-it09}
and to the simplified decoding of intersymbol interference (ISI) \cite{Taghavi-it11}.
In particular, this section describes an extension of the LP decoder
to the joint-decoding of binary-input FSCs and defines LP joint-decoding
pseudo-codewords (JD-PCWs) \cite{Kim-isit10}. This extension is natural
because Feldman's LP formulation of a trellis decoder is general enough
to allow optimal (Viterbi style) decoding of FSCs, and the constraints
associated with the outer LDPC code can be included in the same LP.
This type of extension has been considered as a challenging open problem
in prior works \cite{Feldman-2003}\cite{Wadayama-it10} and was first
given by Flanagan \cite{Flanagan-aller08}\cite{Flanagan-arxiv09},
but was discovered independently by us and reported in \cite{Kim-isit10}.
In particular, Flanagan showed that any communication system which
admits a sum-product (SP) receiver also admits a corresponding linear-programming
(LP) receiver. Since Flanagan's approach is more general, it is also
somewhat more complicated; the resulting LPs are mathematically equivalent
though. One benefit of restricting our attention to FSCs is that our
description of the LP is based on finding a path through a trellis,
which is somewhat more natural for the joint-decoding problem.

These LP decoders provide a natural definition of PCWs for joint-decoding,
and they allow new insight into the joint-decoding problem. Joint-decoding
pseudo-codewords (JD-PCWs) are defined and the decoder error-rate
is upper bounded by a union bound sum over JD-PCWs. This leads naturally
to a provable upper bound (e.g., a union bound) on the probability
of LP decoding failure as a sum over all codewords and JD-PCWs. Moreover,
we can show that all integer solutions are indeed codewords and that
this joint LP decoder also has an ML certificate property. Therefore,
all decoder failures can be explained by (fractional) JD-PCWs. It
is worth noting that this property is not guaranteed by other convex
relaxations of the same problem (e.g., see Wadayama's approach based
on quadratic programming \cite{Wadayama-it10}). %
\begin{comment}
This extension has been considered as a challenging open problem in
the prior works \cite{Wadayama-it10}\cite{Feldman-2003} and the
problem is well posed by Feldman in his PhD thesis \cite[Section 9.5 page 146]{Feldman-2003},
\begin{quote}
\emph{\textquotedbl{}In practice, channels are generally not memoryless
due to physical effects in the communication channel.'' ... {}``Even
coming up with a proper linear cost function for an LP to use in these
channels is an interesting question. The notions of pseudocodeword
and fractional distance would also need to be reconsidered for this
setting.\textquotedbl{} }
\end{quote}
Other than providing satisfying answer to the above open question, 
\end{comment}
{}

Our primary motivation is the prediction of the error rate for joint-decoding
at high SNR. The basic idea is to run simulations at low SNR and keep
track of all observed codeword and pseudo-codeword errors. An estimate
of the error rate at high SNR is computed using a truncated union
bound formed by summing over all observed error patterns at low SNR.
Computing this bound is complicated by the fact that the loss of channel
symmetry implies that the dominant PCWs may depend on the transmitted
sequence. Still, this technique provides a new tool to analyze the
error rate of joint decoders for FSCs and low-density parity-check
(LDPC) codes. Thus, novel prediction results are given in Section
\ref{sec:Simulations}.

\subsection{Joint LP Decoding Derivation}

Now, we describe \emph{the joint LP decoder} in terms of the trellis
of the FSC and the checks in the binary linear code%
\footnote{It is straightforward to extend this joint LP decoder to non-binary
linear codes based on \cite{Flanagan-it09}.%
}. Let $N$ be the length of the code and $\mathbf{y}=(y_{1},y_{2},\ldots,y_{N})$
be the received sequence. The trellis consists of $(N+1)|\mathcal{S}|$
vertices (i.e., one for each state and time) and a set of at most
$2N|\mathcal{S}|^{2}$ edges (i.e., one edge for each input-labeled
state transition and time). The LP formulation requires one indicator
variable for each edge $e\in\mathcal{T}_{i}$, and we denote that
variable by $g_{i,e}$. So, $g_{i,e}$ is equal to 1 if the candidate
path goes through the edge $e$ in $\mathcal{T}_{i}$. Likewise, the
LP decoder requires one cost variable for each edge and we associate
the branch metric $b_{i,e}$ with the edge $e$ given by \[
b_{i,e}\negthinspace\triangleq\negthinspace\begin{cases}
-\ln P\negthinspace\left(y_{t(e)},s'(e)|x(e),s(e)\right)\!\!\! & \mbox{if}\, t(e)\!>\!1\\
-\ln\left[P\left(y_{t(e)},s'(e)|x(e),s(e)\right)\! P_{0}\left(s(e)\right)\right]\!\!\! & \mbox{if}\, t(e)\!=\!1.\end{cases}\]
First, we define the trellis polytope $\mathcal{T}$ formally below.
\begin{defn}
\label{def:trellis-polytope}The \emph{trellis polytope} $\mathcal{T}$
enforces the flow conservation constraints for channel decoder. The
flow constraint for state $k$ at time $i$ is given by\[
\mathcal{F}_{i,k}\triangleq\left\{ \mathbf{g}\in[0,1]^{N\times O}\left|\,\sum_{\substack{e:s'(e)=k}
}g_{i,e}=\sum_{\substack{e:s(e)=k}
}g_{i+1,e}\right.\right\} .\]
Using this, the \emph{trellis polytope} $\mathcal{T}$ is given by\[
\mathcal{T}\triangleq\left\{ \mathbf{g}\in\bigcap_{i=1}^{N-1}\bigcap_{k\in\mathcal{S}}\mathcal{F}_{i,k}\left|\,\sum_{\substack{e\in\mathcal{T}_{p}}
}g_{p,e}=1,\,\mbox{for any}\, p\in\mathcal{I}\right.\right\} .\]

\end{defn}
From simple flow-based arguments, it is known that ML edge path on
trellis can be found by solving a minimum-cost LP applied to the trellis
polytope $\mathcal{T}$.
\begin{thm}
[{\cite[p. 94]{Feldman-2003}}] \label{thm:TLP} \emph{Finding the
ML edge-path through a weighted trellis is equivalent to solving the
minimum-cost flow LP\[
\argmin_{\mathbf{g}\in\mathcal{T}}\sum_{i\in\mathcal{I}}\sum_{\substack{e\in\mathcal{T}_{i}}
}b_{i,e}g_{i,e}\]
and the optimum $\mathbf{g}$ must be integral (i.e., $\mathbf{g}\in\left\{ 0,1\right\} ^{N\times O}$)
unless there are ties.}
\end{thm}
The indicator variables $g_{i,e}$ are used to define the LP and the
code constraints are introduced by defining an auxiliary variable
$f_{i}$ for each code bit.
\begin{defn}
\label{def:Projection}Let the code-space projection\emph{ $\mathcal{Q},$
}be the mapping from $\mathbf{g}$ to the input vector $\mathbf{f}=\left(f_{1},\ldots,f_{N}\right)\in[0,1]^{N}$
defined by $\mathbf{f}=\mathcal{Q}\left(\mathbf{g}\right)$ with \[
f_{i}=\sum_{\substack{e\in\mathcal{T}_{i}:\, x(e)=1}
}g_{i,e}.\]

\end{defn}
For the trellis polytope $\mathcal{T}$, $\mathcal{P}_{\mathcal{T}}(H)$
is the set of vectors whose projection lies inside the relaxed codeword
polytope $\mathcal{P}(H)$.
\begin{defn}
\label{def:TPOLY}The \emph{trellis-wise relaxed polytope} $\mathcal{P}_{\mathcal{T}}(H)$
for $\mathcal{P}(H)$ is given by 

\[
\mathcal{P}_{\mathcal{T}}(H)\triangleq\left\{ \mathbf{g}\in\mathcal{T}\left|\mathcal{Q}\left(\mathbf{g}\right)\in\mathcal{P}(H)\right.\right\} .\]

\end{defn}
The polytope $\mathcal{P}_{\mathcal{T}}(H)$ has integral vertices
which are in one-to-one correspondence with the set of trelliswise
codewords.
\begin{defn}
\label{prop:TCWPOLY}The \emph{set of trellis-wise codewords} $\mathcal{C}_{\mathcal{T}}$
for $\mathcal{C}$ is defined by \[
\mathcal{C}_{\mathcal{T}}\triangleq\left\{ \mathbf{g}\in\mathcal{P}_{\mathcal{T}}(H)\left|\mathbf{g}\in\left\{ 0,1\right\} ^{N\times O}\right.\right\} .\]

\end{defn}
Finally, the joint LP decoder and its ML certificate property are
characterized by the following theorem.
\begin{thm}
%
\begin{comment}
{[}\textbackslash{}cite\{Kim-isit10\}{]}
\end{comment}
{}\label{thm:JointLP}\emph{The LP joint decoder computes\begin{equation}
\argmin_{\mathbf{g}\in\mathcal{P}_{\mathcal{T}}(H)}\sum_{i\in\mathcal{I}}\sum_{\substack{e\in\mathcal{T}_{i}}
}b_{i,e}g_{i,e}\label{eq:JointLP}\end{equation}
and outputs a joint ML edge-path if $\mathbf{g}$ is integral.}\end{thm}
\begin{IEEEproof}
Let $\mathcal{V}$ be the set of valid input/state sequence pairs.
For a given $\mathbf{y}$, the ML edge-path decoder finds the most
likely path, through the channel trellis, whose input sequence is
a codeword. Mathematically, it computes \begin{align*}
 & \!\!\!\argmax_{(x_{1}^{N},s_{0}^{N})\in\mathcal{V}}P(y_{1}^{N},s_{1}^{N}|x_{1}^{N},s_{0})P_{0}\left(s(e)\right)\\
 & =\argmax_{\mathbf{g}\in\mathcal{C}_{\mathcal{T}}}P_{0}\left(s(e)\right)\prod_{i\in\mathcal{I}}\prod_{\substack{\substack{e\in\mathcal{T}_{i}:\, g_{i,e}=1}
}
}\negthinspace\!\!\!\!\negthinspace\negthinspace P\negthinspace\left(y_{t(e)},s'(e)|x(e),s(e)\right)\\
 & =\argmin_{\mathbf{g}\in\mathcal{C}_{\mathcal{T}}}\sum_{i\in\mathcal{I}}\sum_{\substack{e\in\mathcal{T}_{i}:\, g_{i,e}=1}
}b_{i,e}\\
 & =\argmin_{\mathbf{g}\in\mathcal{C}_{\mathcal{T}}}\sum_{i\in\mathcal{I}}\sum_{\substack{e\in\mathcal{T}_{i}}
}b_{i,e}g_{i,e},\end{align*}
where ties are resolved in a systematic manner and $b_{1,e}$ has
the extra term $-\ln\, P_{0}\left(s(e)\right)$ for the initial state
probability. By relaxing $\mathcal{C}_{\mathcal{T}}$ into $\mathcal{P}_{\mathcal{T}}(H)$,
we obtain the desired result.\end{IEEEproof}
\begin{cor}
%
\begin{comment}
{[}\textbackslash{}cite\{Kim-isit10\}{]}
\end{comment}
{}\label{cor:JointLPISI}\emph{For a FSISIC}%
\footnote{In fact, this holds more generally for the restricted class of FSCs
used in \cite{Ziv-it85}, which are now called unifilar FSCs because
they generalize the unifilar Markov sources defined in \cite{Ash-1990}.%
}\emph{, the LP joint decoder outputs a joint ML codeword if $\mathbf{g}$
is integral.}\end{cor}
\begin{IEEEproof}
The joint ML decoder for codewords computes\begin{align*}
 & \!\!\!\argmax_{x_{1}^{N}\in\mathcal{C}}\sum_{s_{1}^{N}\in\mathcal{S}^{N}}P(y_{1}^{N},s_{1}^{N}|x_{1}^{N},s_{0})P_{0}\left(s(e)\right)\\
 & =\argmax_{x_{1}^{N}\in\mathcal{C}}\sum_{s_{1}^{N}\in\mathcal{S}^{N}}\prod_{i\in\mathcal{I}}P(y_{i},s_{i+1}|x_{i},s_{i})P_{0}\left(s(e)\right)\\
 & \stackrel{(a)}{=}\argmax_{x_{1}^{N}\in\mathcal{C}}\prod_{i\in\mathcal{I}}P\left(y_{i},\eta\left(x_{i},s_{i}\right)\big|x_{i},s_{i}\right)P_{0}\left(s(e)\right)\\
 & \stackrel{(b)}{=}\argmin_{\mathbf{g}\in\mathcal{C}_{\mathcal{T}}}\sum_{i\in\mathcal{I}}\sum_{\substack{e\in\mathcal{T}_{i}}
}b_{i,e}g_{i,e},\end{align*}
where $(a)$ follows from Definition \ref{def:FSISI} and $(b)$ holds
because each input sequence defines a unique edge-path. Therefore,
the LP joint-decoder outputs an ML codeword if $\mathbf{g}$ is integral.\end{IEEEproof}
\begin{rem}
\label{rem:MLedgepath} If the channel is not a FSISIC (e.g., if it
is a finite-state fading channel), then integer valued solutions of
the LP joint-decoder are ML edge-paths but not necessarily ML codewords.
This occurs because the joint LP decoder does not sum the probability
of the multiple edge-paths associated with the same codeword (e.g.,
when multiple distinct edge-paths are associated with the same input
labels). Instead, it simply gives the probability of the most-likely
edge path associated that codeword.
\end{rem}
\begin{figure}[t]
\begin{centering}
\includegraphics[scale=0.65]{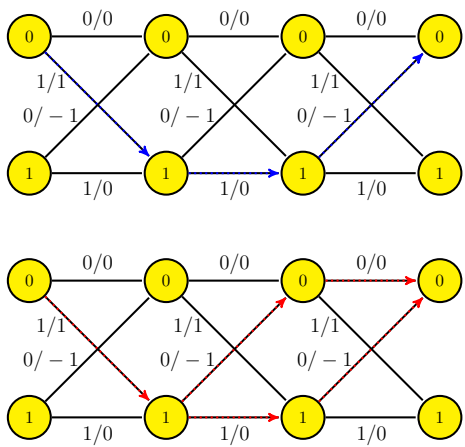}
\par\end{centering}

\caption{\label{fig:Example}Illustration of joint LP decoder outputs for the
single parity-check code SPC(3,2) over DIC (starts in zero state).
By ordering the trellis edges appropriately, joint LP decoder converges
to either a TCW $(0\,1\,0\,0;0\,0\,0\,1;.0\,0\,1\,0)$ (top dashed
blue path) or a JD-TPCW $(0\,1\,0\,0;0\,0\,.5\,.5;.5\,0\,.5\,0)$
(bottom dashed red paths). Using $\mathcal{Q}$ to project them into
$\mathcal{P}(H)$, we obtain the corresponding SCW $(1,1,0)$ and
JD-SPCW $(1,.5,0).$ }

\end{figure}

\subsection{Joint LP Decoding Pseudo-codewords \label{sec:JD-PCW}}

Pseudo-codewords have been observed and given names by a number of
authors (e.g., \cite{Wiberg-96,Di-it02,Richardson-aller03}), but
the simplest general definition was provided by Feldman et al. in
the context of LP decoding of parity-check codes \cite{Feldman-it05}.
One nice property of the LP decoder is that it always returns either
an integral codeword or a fractional pseudo-codeword. Vontobel and
Koetter have shown that a very similar set of pseudo-codewords also
affect message-passing decoders, and that they are essentially fractional
codewords that cannot be distinguished from codewords using only local
constraints \cite{Vontobel-arxiv05}. The joint-decoding pseudo-codeword
(JD-PCW), defined below, can be used to characterize code performance
at low error rates. 
\begin{defn}
\label{def:TCW}If $g_{i,e}\in\{0,\,1\}$ for all $e$, then the output
of the LP joint decoder is a\textbf{ }\emph{trellis-wise }%
\begin{comment}
\emph{(ML) }
\end{comment}
{}\emph{codeword} (TCW). Otherwise, $g_{i,e}\in(0,\,1)$ for some $e$
and the solution is called a \emph{joint-decoding trellis-wise pseudo-codeword}
(JD-TPCW); in this case, the decoder outputs {}``failure'' (see
Fig. \ref{fig:Example} for an example of this definition). 
\end{defn}

\begin{defn}
\label{def:SCW}For any TCW $\mathbf{g}$, the projection $\mathbf{f}=\mathcal{Q}\left(\mathbf{g}\right)$
is called a \emph{symbol-wise codeword} (SCW). Likewise, for any JD-TPCW
$\mathbf{g}$, the projection $\mathbf{f}=\mathcal{Q}\left(\mathbf{g}\right)$
is called a \emph{joint-decoding symbolwise pseudo-codeword} (JD-SPCW)
(see Fig. \ref{fig:Example} for a graphical depiction of this definition).
%
\begin{comment}
in many-to-one manner (depends on the initial state assumption)
\end{comment}
{}%
\begin{comment}
in one-to-many manner for $\mathbf{f}=\left(f_{1},\, f_{2},\,\ldots\,,\, f_{n}\right)$
where
\end{comment}
{}\end{defn}
\begin{rem}
\label{rem:MLcertificate}For FSISICs, the LP joint decoder has the
\emph{ML certificate }property; if the decoder outputs a SCW, then
it is guaranteed to be the ML codeword (see Corollary \ref{cor:JointLPISI}).
\end{rem}

\begin{defn}
If $\mathbf{g}$ is a JD-TPCW, then $\mathbf{p}=\left(p_{1},\ldots,p_{N}\right)$
with \[
p_{i}=\sum_{\substack{e\in\mathcal{T}_{i}}
}g_{i,e}a\left(e\right),\]
is called a \emph{joint-decoding symbol-wise signal-space pseudo-codeword}
(JD-SSPCW). Likewise, if $\mathbf{g}$ is a TCW, then $\mathbf{p}$
is called a \emph{symbol-wise signal-space codeword} (SSCW).%
\begin{comment}
Any TCW can be projected onto a \emph{symbol-wise signal-space codeword}
(SSCW) in many-to-one manner (depends on the initial state assumption)
and any JD-TPCW $\mathbf{g}$ can be projected onto a\emph{ joint-decoding
symbol-wise signal-space pseudo-codeword} (JD-SSPCW) $\mathbf{p}=\left(p_{1},\ldots,p_{N}\right)$
by averaging the components with \[
p_{i}=\sum_{\substack{e\in\mathcal{T}_{i}}
}g_{i,e}a\left(e\right).\]

\end{comment}
{}%
\begin{comment}
\begin{example}
Consider the single parity-check code SPC(3,2). Over precoded dicode
channel (starts in zero state) {[}change to DIC??{]} with AWGN, this
code has five joint-decoding pseudo-codewords. A simulation was performed
for joint-decoding of the SPC(3,2) on the pDIC trellis and the set
of JD-TPCW, by ordering the trellis edges appropriately, was found
to be \begin{align*}
\{(0\,1\,0\,0;0\,0\,.5\,.5;0\,.5\,.5\,0), & (.5\,.5\,0\,0;.5\,0\,0\,.5;0\,1\,0\,0),\\
(.5\,.5\,0\,0;0\,.5\,.5\,0;0\,0\,1\,0), & (1\,0\,0\,0;.5\,.5\,0\,0;0\,.5\,.5\,0),\\
(.5\,.5\,0\,0;.5\,0\,0\,0;0\,.5\,.5\,0)\}.\end{align*}
Using $\mathcal{Q}$ to project them into $\mathcal{P}(H)$, we get
the corresponding set of JD-SPCW \[
\{(1,.5,.5),\,(.5,.5,1),\,(.5,.5,0),\,(0,.5,.5),\,(.5,0,.5)\}.\]

\end{example}

\end{comment}
{}
\end{defn}

\subsection{Union Bound for Joint LP Decoding \label{sec:Bound}}

Now that we have defined the relevant pseudo-codewords, we consider
how much a particular pseudo-codeword affects performance; the idea
is to quantify pairwise error probabilities. In fact, we will use
the insights gained in the previous section to obtain a union bound
on the decoder's word-error probability%
\begin{comment}
(as a tight approximation)
\end{comment}
{} and to analyze the performance of the proposed joint LP decoder.
Toward this end, let's consider the pairwise error event between a
SSCW $\mathbf{c}$ and a JD-SSPCW $\mathbf{p}$ first.
\begin{thm}
\label{thm:PEE}\emph{A necessary and sufficient condition for the
pairwise decoding error between a SSCW $\mathbf{c}$ and a JD-SSPCW
$\mathbf{p}$ is} \begin{equation}
\sum_{i\in\mathcal{I}}\sum_{\substack{e\in\mathcal{T}_{i}}
}b_{i,e}g_{i,e}\leq\sum_{i\in\mathcal{I}}\sum_{\substack{e\in\mathcal{T}_{i}}
}b_{i,e}\tilde{g}_{i,e},\label{eq:error condition}\end{equation}
\emph{where $\mathbf{g}\in\mathcal{P}_{\mathcal{T}}(H)$ and $\tilde{\mathbf{g}}\in\mathcal{C}_{\mathcal{T}}$
are the LP variables for $\mathbf{p}$ and $\mathbf{c}$ respectively. }\end{thm}
\begin{IEEEproof}
By definition, the joint LP decoder \eqref{eq:JointLP} prefers \emph{$\mathbf{p}$
}over \emph{$\mathbf{c}$ }if and only if \eqref{eq:error condition}
holds.
\end{IEEEproof}
%
\begin{comment}
Since the condition in Theorem. \ref{thm:PEE} is difficult to handle
directly, we are currently working on to obtain the general expression
that breaks the error probability into PCW of the code and error events
in the channel trellis. 
\end{comment}
{}For the moment, let $\mathbf{c}$ be the SSCW of FSISIC to an AWGN
channel whose output sequence is $\mathbf{y}=\mathbf{c}+\mathbf{v}$,
where $\mathbf{v}=(v_{1},\ldots,v_{N})$ is an i.i.d. Gaussian sequence
with mean $0$ and variance $\sigma^{2}$. Then, the joint LP decoder
can be simplified as stated in the Theorem \ref{thm:LPAWGN}.
\begin{thm}
\label{thm:LPAWGN} \emph{Let $\mathbf{y}$ be the output of a FSISIC
with zero-mean AWGN whose variance is $\sigma^{2}$ per output. Then,
the joint LP decoder is equivalent to} \[
\argmin_{\mathbf{g}\in\mathcal{P}_{\mathcal{T}}(H)}\sum_{i\in\mathcal{I}}\sum_{\substack{e\in\mathcal{T}_{i}}
}\left(y_{i}-a\left(e\right)\right)^{2}g_{i,e}.\]
\end{thm}
\begin{IEEEproof}
For each edge $e$, the output $y_{i}$ is Gaussian with mean $a\left(e\right)$
and variance $\sigma^{2}$, so we have $P\left(y_{t(e)},s'(e)|x(e),s(e)\right)\sim\mathcal{N}\left(a\left(e\right),\,\sigma^{2}\right)$.
Therefore, the joint LP decoder computes\[
\argmin_{\mathbf{g}\in\mathcal{P}_{\mathcal{T}}(H)}\sum_{i\in\mathcal{I}}\sum_{\substack{e\in\mathcal{T}_{i}}
}b_{i,e}g_{i,e}=\argmin_{\mathbf{g}\in\mathcal{P}_{\mathcal{T}}(H)}\sum_{i\in\mathcal{I}}\sum_{\substack{e\in\mathcal{T}_{i}}
}\left(y_{i}-a\left(e\right)\right)^{2}g_{i,e}.\]

\end{IEEEproof}
%
\begin{comment}
\begin{IEEEproof}
\begin{align*}
 & \argmin_{g(\cdot)\in\mathcal{P}_{\mathcal{T}}(H)}\left(-\sum_{e\in\mathcal{E}}g(e)\,\ln\, P\left(y_{t(e)},s'(e)|x(e),s(e)\right)\right)\\
 & =\argmin_{g(\cdot)\in\mathcal{P}_{\mathcal{T}}(H)}\sum_{e\in\mathcal{E}}g(e)\left(y_{t(e)}-a(e)\right)^{2}.\end{align*}

\end{IEEEproof}

\end{comment}
{}We will show that each pairwise probability has a simple closed-form
expression that depends only on a generalized squared Euclidean distance
$d_{gen}^{2}\left(\mathbf{c},\,\mathbf{p}\right)$ and the noise variance
$\sigma^{2}.$ One might notice that this result is very similar to
the pairwise error probability derived in \cite{Forney-ima99}. The
main difference is the trellis-based approach that allows one to obtain
this result for FSCs. Therefore, the next definition and theorem can
be seen as a generalization of \cite{Forney-ima99}%
\begin{comment}
 and a special case of the more general formulation in \cite{Flanagan-arxiv09}
\end{comment}
{}.
\begin{defn}
\label{def:dist}Let $\mathbf{c}$ be a SSCW and $\mathbf{p}$ a JD-SSPCW.
Then the \emph{generalized squared Euclidean distance} between $\mathbf{c}$
and $\mathbf{p}$ can be defined in terms of their trellis-wise descriptions
by \[
d_{gen}^{2}\left(\mathbf{c},\,\mathbf{p}\right)\triangleq\frac{\left(\left\Vert \mathbf{d}\right\Vert ^{2}+\sigma_{p}^{2}\right)^{2}}{\left\Vert \mathbf{d}\right\Vert ^{2}}\]
where

\vspace{-1mm}

\begin{align*}
\left\Vert \mathbf{d}\right\Vert ^{2} & \triangleq\sum_{i\in\mathcal{I}}\left(c_{i}-p_{i}\right)^{2},\,\sigma_{p}^{2}\triangleq\sum_{i\in\mathcal{I}}\sum_{\substack{e\in\mathcal{T}_{i}}
}g_{i,e}a^{2}\left(e\right)-\sum_{i\in\mathcal{I}}p_{i}^{2}.\end{align*}

\end{defn}

\begin{thm}
\label{thm:PEP}\emph{The pairwise error probability between a SSCW
$\mathbf{c}$ and a JD-SSPCW} $\mathbf{p}$ \emph{is\[
\mbox{Pr}\left(\mathbf{c}\rightarrow\mathbf{p}\right)=Q\left(\frac{d_{gen}\left(\mathbf{c},\,\mathbf{p}\right)}{2\sigma}\right),\]
where $Q\left(x\right)=\frac{1}{\sqrt{2\pi}}\int_{x}^{\infty}\texttt{e}^{-t^{2}/2}dt$}%
\begin{comment}
\emph{as usual the integral from $\theta$ to $\infty$ of the Gaussian
distribution with mean 0 and variance 1}
\end{comment}
{}\emph{.}\end{thm}
\begin{IEEEproof}
The pairwise error probability $\mbox{Pr}\left(\mathbf{c}\rightarrow\mathbf{p}\right)$
that the LP joint-decoder will choose the pseudo-codeword $\mathbf{p}$
over $\mathbf{c}$ can be written as \begin{align*}
 & \!\!\!\mbox{Pr}\left(\mathbf{c}\rightarrow\mathbf{p}\right)\\
 & =\mbox{Pr}\left\{ \sum_{i\in\mathcal{I}}\sum_{\substack{e\in\mathcal{T}_{i}}
}g_{i,e}\left(y_{i}-a\left(e\right)\right)^{2}\leq\sum_{i\in\mathcal{I}}\left(y_{i}-c_{i}\right)^{2}\right\} \\
 & =\mbox{Pr}\left\{ \begin{array}{c}
\sum_{i}y_{i}\left(c_{i}-p_{i}\right)\leq\frac{1}{2}\left(\sum_{i}c_{i}^{2}-\sum_{i}\sum_{\substack{e}
}g_{i,e}a^{2}\left(e\right)\right)\end{array}\right\} \\
 & \stackrel{(a)}{=}Q\left(\frac{\sum_{i}c_{i}\left(c_{i}-p_{i}\right)-\frac{1}{2}\left(\sum_{i}c_{i}^{2}-\sum_{i}\sum_{\substack{e}
}g_{i,e}a^{2}\left(e\right)\right)}{\sigma\sqrt{\sum_{i}\left(c_{i}-p_{i}\right)^{2}}}\right)\\
 & \stackrel{(b)}{=}Q\left(\frac{\left\Vert \mathbf{d}\right\Vert ^{2}+\sigma_{p}^{2}}{2\sigma\left\Vert \mathbf{d}\right\Vert }\right)=Q\left(\frac{d_{gen}\left(\mathbf{c},\,\mathbf{p}\right)}{2\sigma}\right),\end{align*}
where $(a)$ follows from the fact that $\sum_{i}y_{i}\left(c_{i}-p_{i}\right)$
has a Gaussian distribution with mean $\sum_{i}c_{i}(c_{i}-p_{i})$
and variance $\sum_{i}(c_{i}-p_{i})^{2}$, and $(b)$ follows from
Definition \ref{def:dist}. 
\end{IEEEproof}
The performance degradation of LP decoding relative to ML decoding
can be explained by pseudo-codewords and their contribution to the
error rate, which depends on $d_{gen}\left(\mathbf{c},\,\mathbf{p}\right).$
Indeed, by defining $K_{d_{gen}}(\mathbf{c})$ as the number of codewords
and JD-PCWs at distance $d_{gen}$ from $\mathbf{c}$ and $\mathcal{G}(\mathbf{c})$
as the set of generalized Euclidean distances, we can write the union
bound on word error rate (WER) as \begin{equation}
P_{w|\mathbf{c}}\leq\sum_{d_{gen}\in\mathcal{G}(\mathbf{c})}K_{d_{gen}}(\mathbf{c})\, Q\left(\frac{d_{gen}}{2\sigma}\right).\label{eq:UnionBound}\end{equation}
Of course, we need the set of JD-TPCWs to compute $\mbox{Pr}\left(\mathbf{c}\rightarrow\mathbf{p}\right)$
with the Theorem \ref{thm:PEP}. There are two complications with
this approach. One is that, like the original problem \cite{Feldman-2003},
no general method is known yet for computing the generalized Euclidean
distance spectrum efficiently%
\begin{comment}
, apart from going through all error events explicitly
\end{comment}
{}. Another is, unlike original problem, the constraint polytope may
not be symmetric under codeword exchange. Therefore the decoder performance
may not be symmetric under codeword exchange. Hence, the decoder performance
may depend on the transmitted codeword. In this case, the pseudo-codewords
will also depend on the transmitted sequence. %
\begin{comment}
%
\begin{figure}
%
\framebox{\begin{minipage}[c]{0.97\columnwidth}%
\selectlanguage{english}%
\vspace{2mm}

\selectlanguage{american}%
\textbf{Problem-P:}\\
%
\begin{minipage}[t]{1\columnwidth}%
\vspace{-7mm}\[
\min_{\mathbf{g,w}}\sum_{i\in\mathcal{I}}\sum_{e\in\mathcal{T}_{i}}b_{i,e}g_{i,e}\]

\selectlanguage{english}%
\vspace{-5mm}

\selectlanguage{american}%
subject to\[
\sum_{\mathcal{B}\in\mathcal{E}_{j}}w_{j,\mathcal{B}}=1,\,\,\,\forall j\in\mathcal{J},\,\,\,\sum_{\substack{e\in\mathcal{T}_{p}}
}g_{p,e}=1,\,\mbox{for any}\, p\in\mathcal{I}\]
\[
\sum_{\mathcal{B}\in\mathcal{E}_{j},\mathcal{B}\ni i}w_{j,\mathcal{B}}=\sum_{e:x(e)=1}g_{i,e},\,\,\,\forall i\in\mathcal{I},j\in\mathcal{N}\left(i\right)\]

\[
\sum_{e:s'(e)=k}g_{i,e}=\sum_{e:s(e)=k}g_{i+1,e},\,\,\,\forall i\in\mathcal{I}\setminus N,\, k\in\emph{S}\]
\[
w_{j,\mathcal{B}}\geq0,\,\,\,\forall j\in\mathcal{J},\,\mathcal{B}\in\mathcal{E}_{j},\,\,\, g_{i,e}\geq0,\,\,\,\forall i\in\mathcal{I},\, e\in\mathcal{T}_{i}.\]

\vspace{-1mm}%
\end{minipage}%
\end{minipage}}
\end{figure}

\end{comment}
{}

\section{Iterative Solver for the Joint LP Decoder \label{sec:IterativeJoint-LPD}}

In the past, the primary value of linear programming (LP) decoding
was as an analytical tool that allowed one to better understand iterative
decoding and its modes of failure. This is because LP decoding based
on standard LP solvers is quite impractical and has a superlinear
complexity in the block length. This motivated several authors to
propose low-complexity algorithms for LP decoding of LDPC codes in
the last five years (e.g., \cite{Wadayama-it10,Vontobel-turbo06,Vontobel-ita08,Taghavi-it08,Wadayama-isit09,Burshtein-it09,Punekar-aller10}).
Many of these have their roots in the iterative Gauss-Seidel approach
proposed by Vontobel and Koetter for approximate LP decoding \cite{Vontobel-turbo06}.
This approach was also analyzed further by Burshtein \cite{Burshtein-it09}.
Smoothed Lagrangian relaxation methods have also been proposed to
solve intractable optimal inference and estimation for more general
graphs (e.g., \cite{Johnson-2008}). 

In this section, we consider the natural extension of \cite{Vontobel-turbo06}\cite{Burshtein-it09}
to the joint-decoding LP formulation developed in Section \ref{sec:Joint-LPD}%
\begin{comment}
 in Section \ref{sub:Iterative-Joint-LP}
\end{comment}
{}. We argue that, by taking advantage of the special dual-domain structure
of the joint LP problem and replacing minima in the formulation with
soft-minima, we can obtain an efficient method that solves the joint
LP. While there are many ways to iteratively solve the joint LP, our
main goal was to derive one as the natural analogue of turbo equalization
(TE). This should lead to an efficient method for joint LP decoding
whose performance is similar to that of joint LP and whose per-iteration
complexity similar to that of TE. Indeed, the solution we provide
is a fast, iterative, and provably convergent form of TE whose update
rules are tightly connected to BCJR-based TE. This demonstrates that
an iterative joint LP solver with a similar computational complexity
as TE is feasible (see Remark \ref{rem:TEconnection}). In practice,
the complexity reduction of this iterative decoder comes at the expense
of some performance loss, when compared to the joint LP decoder, due
to convergence issues (discussed in Section \ref{sub:Convergence}).

Previously, a number of authors have attempted to reverse engineer
an objective function targeted by turbo decoding (and TE by association)
in order to discuss its convergence and optimality \cite{Regalia-07,Alberge-eusipco08,Walsh-it10}.
For example, \cite{Regalia-07} uses a duality link between two optimality
formulations of TE: one based on Bethe free energy optimization and
the other based on constrained ML estimation. This results of this
section establish a new connection between iterative decoding and
optimization for the joint-decoding problem that can also be extended
to turbo decoding.

%
\begin{comment}
%
\begin{figure}
%
\framebox{\begin{minipage}[c]{0.97\columnwidth}%
\selectlanguage{english}%
\vspace{2mm}

\selectlanguage{american}%
\textbf{Problem-D1:}\\
%
\begin{minipage}[t]{1\columnwidth}%
\selectlanguage{english}%
\vspace{-5mm}\foreignlanguage{american}{\[
\max_{\mathbf{m,n}}\sum_{j\in\emph{J}}\min_{\mathcal{B}\in\mathcal{E}_{j}}\!\left[\sum_{i\in\mathcal{B}}m_{i,j}\right]\!\!+\!\min_{e\in\mathcal{T}_{p}}\!\left[\Gamma_{p,e}\!-\! n_{p-1,s(e)}\!+\! n_{p,s'(e)}\right]\]
subject to \vspace{-0mm}\[
\Gamma_{i,e}\geq n_{i-1,s(e)}-n_{i,s'(e)},\,\forall i\in\mathcal{I}\setminus p,\, e\in\mathcal{T}_{i}\vspace{-2mm}\]
and \vspace{-4mm}}

\selectlanguage{american}%
\[
n_{0,k}=n_{N,k}=0,\,\forall k\in\emph{S},\vspace{-2mm}\]

where \vspace{-4mm}

\emph{\[
\Gamma_{i,e}\triangleq b_{i,e}-\delta_{x(e)=1}\sum_{j\in\mathcal{N}(i)}m_{i,j}.\]
}%
\end{minipage}%
\end{minipage}}
\end{figure}

\end{comment}
{}%
\begin{table}[b]
\captionsetup{name=Table} \caption{Primal Problem (Problem-P)}\label{Flo:primal}\[
\min_{\mathbf{g,w}}\sum_{i\in\mathcal{I}}\sum_{e\in\mathcal{T}_{i}}b_{i,e}g_{i,e}\]

\selectlanguage{english}%
\vspace{-5mm}

\selectlanguage{american}%
subject to\[
\sum_{\mathcal{B}\in\mathcal{E}_{j}}w_{j,\mathcal{B}}=1,\,\,\,\forall j\in\mathcal{J},\,\,\,\sum_{\substack{e\in\mathcal{T}_{p}}
}g_{p,e}=1,\,\mbox{for any}\, p\in\mathcal{I}\]
\[
\sum_{\mathcal{B}\in\mathcal{E}_{j},\mathcal{B}\ni i}w_{j,\mathcal{B}}=\sum_{e:x(e)=1}g_{i,e},\,\,\,\forall i\in\mathcal{I},j\in\mathcal{N}\left(i\right)\]

\[
\sum_{e:s'(e)=k}g_{i,e}=\sum_{e:s(e)=k}g_{i+1,e},\,\,\,\forall i\in\mathcal{I}\setminus N,\, k\in\emph{S}\]
\[
w_{j,\mathcal{B}}\geq0,\,\,\,\forall j\in\mathcal{J},\,\mathcal{B}\in\mathcal{E}_{j},\,\,\, g_{i,e}\geq0,\,\,\,\forall i\in\mathcal{I},\, e\in\mathcal{T}_{i}.\]

\end{table}

\subsection{Iterative Joint LP Decoding Derivation \label{sub:Iterative-Joint-LP}}

\begin{figure}[t]
\begin{centering}
\includegraphics[scale=0.65]{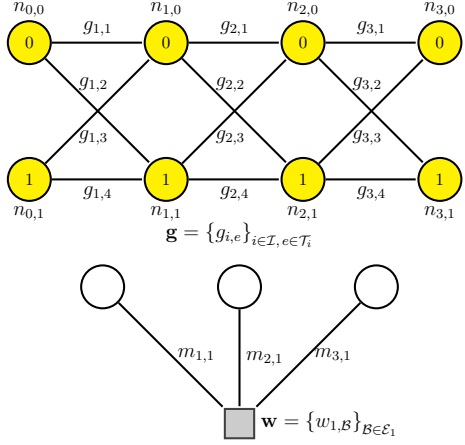}
\par\end{centering}

\caption{\label{fig:ProblemP}Illustration of primal variables $\mathbf{g}$
and $\mathbf{w}$ defined for Problem-P and dual variables $\mathbf{n}$
and $\mathbf{m}$ defined for Problem-D1 on the same example given
by Fig. \ref{fig:Example}: SPC(3,2) with DIC for $N=3.$}

\end{figure}

\begin{table}[b]
\captionsetup{name=Table} \caption{Dual Problem 1st Formulation (Problem-D1)}\label{Flo:dual1}\[
\max_{\mathbf{m,n}}\sum_{j\in\emph{J}}\min_{\mathcal{B}\in\mathcal{E}_{j}}\!\left[\sum_{i\in\mathcal{B}}m_{i,j}\right]\!\!+\!\min_{e\in\mathcal{T}_{p}}\!\left[\Gamma_{p,e}\!-\! n_{p-1,s(e)}\!+\! n_{p,s'(e)}\right]\]
subject to \vspace{-0mm}\[
\Gamma_{i,e}\geq n_{i-1,s(e)}-n_{i,s'(e)},\,\forall i\in\mathcal{I}\setminus p,\, e\in\mathcal{T}_{i}\vspace{-2mm}\]
and \vspace{-4mm}

\[
n_{0,k}=n_{N,k}=0,\,\forall k\in\emph{S},\vspace{-2mm}\]

where \vspace{-4mm}

\emph{\[
\Gamma_{i,e}\triangleq b_{i,e}-\delta_{x(e)=1}\sum_{j\in\mathcal{N}(i)}m_{i,j}.\]
}
\end{table}

In Section \ref{sec:Joint-LPD}, joint LP decoder is presented as
an LDPC-code constrained shortest-path problem on the channel trellis.
In this section, we develop the iterative solver for the joint-decoding
LP. There are few key steps in deriving iterative solution for the
joint LP decoding problem. For the first step, given by the primal
problem (Problem-P) in Table \ref{Flo:primal}, we reformulate the
original LP \eqref{eq:JointLP} in Theorem \ref{thm:JointLP} using
only equality constraints involving the indicator variables%
\footnote{The valid patterns $\mathcal{E}_{j}\triangleq\left\{ \mathcal{B}\subseteq\mathcal{N}\left(j\right)|\left|\mathcal{B}\right|\,\mbox{is}\,\mbox{even}\right\} $
for each parity-check $j\in\mathcal{J}$ allow us to define the indicator
variables $w_{j,\mathcal{B}}$ (for $j\in\mathcal{J}$ and $\mathcal{B}\in\mathcal{E}_{j}$)
which equal 1 if the codeword satisfies parity-check $j$ using configuration
$\mathcal{B}\in\mathcal{E}_{j}.$%
} $\mathbf{g}$ and $\mathbf{w}$. The second step, given by the 1st
formulation of the dual problem (Problem-D1) in Table \ref{Flo:dual1},
follows from standard convex analysis (e.g., see \cite[p. 224]{Boyd-2004}).
Strong duality holds because the primal problem is feasible and bounded.
Therefore, the Lagrangian dual of Problem-P is equivalent to Problem-D1
and the minimum of Problem-P is equal to the maximum of Problem-D1.
From now on, we consider Problem-D1, where the code and trellis constraints
separate into two terms in the objective function. See Fig. \ref{fig:ProblemP}
for a diagram of the variables involved. 

\begin{figure}[t]
\begin{centering}
\includegraphics[scale=0.65]{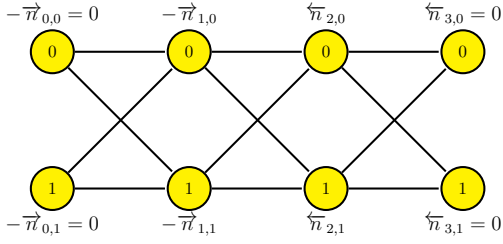}
\par\end{centering}

\caption{\label{fig:ProblemD-2}Illustration of Viterbi updates in Problem-D2
on the same example given by Fig. \ref{fig:Example}: DIC for $N=3$
with forward $\protect\overrightarrow{\mathbf{n}}$ and backward $\protect\overleftarrow{\mathbf{n}}.$ }

\end{figure}

The third step, given by the 2nd formulation of the dual problem (Problem-D2)
in Table \ref{Flo:dual2}, observes that forward/backward recursions
can be used to perform the optimization over $\mathbf{n}$ and remove
one of the dual variable vectors. This splitting is enabled by imposing
the trellis flow normalization constraint in Problem-P only at one
time instant $p\in\mathcal{I}$. This detail gives $N$ different
ways to write the same LP and is an important part of obtaining update
equations similar to those of TE.%
\begin{comment}
%
\begin{figure}
%
\framebox{\begin{minipage}[c]{0.97\columnwidth}%
\selectlanguage{english}%
\vspace{2mm}

\selectlanguage{american}%
\textbf{Problem-D2:}\\
%
\begin{minipage}[t]{1\columnwidth}%
\selectlanguage{english}%
\vspace{-5mm}\foreignlanguage{american}{\[
\max_{\mathbf{m}}\sum_{j\in\emph{J}}\min_{\mathcal{B}\in\mathcal{E}_{j}}\!\left[\sum_{i\in\mathcal{B}}m_{i,j}\right]\!+\min_{e\in\mathcal{T}_{p}}\!\left[\Gamma_{p,e}\!-\!\overrightarrow{n}_{p\!-\!1,s(e)}\!+\!\overleftarrow{n}_{p,s'(e)}\right]\]
where $\overrightarrow{n}_{i,k}$ is defined for $i=1,\,\ldots\,,\, p-1$
by \vspace{0mm} \[
-\overrightarrow{n}_{i,k}=\min_{e\in s'^{-1}(k)}-\overrightarrow{n}_{i-1,s(e_{i})}+\Gamma_{i,e},\,\forall k\in\mathcal{S}\vspace{-1mm}\]
and $\overleftarrow{n}_{i,k}$ is defined for $i=N-1,N-2,\,\ldots\,,\, p$
by \vspace{0mm}\emph{\[
\overleftarrow{n}_{i,k}=\min_{e\in s^{-1}(k)}\overleftarrow{n}_{i+1,s'(e_{i+1})}+\Gamma_{i+1,e},\,\forall k\in\mathcal{S}\vspace{-1.5mm}\]
}starting from \vspace{-1mm}\[
\overrightarrow{n}_{0,k}=\overleftarrow{n}_{N,k}=0,\,\forall k\in\mathcal{S}.\]
 \vspace{-5mm}}\selectlanguage{american}
%
\end{minipage}%
\end{minipage}}
\end{figure}

\end{comment}
{}
\begin{lem}
\emph{\label{lem:Problem-D-2}Problem-D1 is equivalent to Problem-D2.}\end{lem}
\begin{IEEEproof}
By rewriting the inequality constraint in Problem-D1 as\[
-n_{i,s'(e_{i})}\leq-n_{i-1,s(e_{i})}+\Gamma_{i,e}\]
we obtain the recursive upper bound for $i=p-1$ as\begin{align*}
 & -n_{p-1,k}\\
 & \leq\negthinspace\left.-n_{p-2,s(e_{p-1})}+\Gamma_{p-1,e}\right|_{s'(e_{p\!-\!1})=k}\\
 & \leq\negthinspace\left.-n_{p-3,s(e_{p-2})}\negthinspace+\negthinspace\Gamma_{p-2,e}\right|_{s'(e_{p\!-\!2})=s(e_{p\!-\!1})}\negthinspace+\negthinspace\left.\Gamma_{p-1,e}\right|_{s'(e_{p\!-\!1})=k}\\
 & \,\,\,\,\,\,\,\,\,\,\,\,\,\,\,\,\,\,\,\,\,\,\,\,\,\,\,\,\,\,\,\,\,\,\,\,\,\,\,\,\,\,\,\,\,\,\,\,\,\,\,\,\,\,\,\,\vdots\,\,\,\,\,\,\,\,\,\,\,\,\,\,\,\,\,\,\,\,\,\,\,\,\,\,\,\,\,\,\,\,\,\,\\
 & \leq\negthinspace\left.-n_{1,s(e_{2})}\negthinspace+\negthinspace\sum_{i=2}^{p-1}\Gamma_{i,e_{}}\right|_{s'(e_{p\!-\!1})=k,s'(e_{p\!-\!2})=s(e_{p\!-\!1}),\ldots,s'(e_{1})=s(e_{2}).}\end{align*}
This upper bound $-n_{p-1,k}\leq-\overrightarrow{n}_{p-1,k}$ is achieved
by the forward Viterbi update in Problem-D2 for $i=1,\,\ldots\,,\, p-1.$
Again, by expressing the same constraint as \[
n_{i-1,s(e_{i})}\leq\Gamma_{i,e}+n_{i,s'(e_{i})}\]
we get a recursive upper bound for $i=p+1$. Similar reasoning shows
this upper bound $n_{p,k}\leq\overleftarrow{n}_{p,k}$ is achieved
by the backward Viterbi update in Problem-D2 for $i=N-1,N-2,\,\ldots\,,\, p.$
See Fig. \ref{fig:ProblemD-2} for a graphical depiction of this.
\end{IEEEproof}
The fourth step, given by the softened dual problem (Problem-DS) in
Table \ref{Flo:softdual}, is formulated by replacing the minimum
operator in Problem-D2 with the soft-minimum operation \[
\mbox{min}\left(x_{1},\, x_{2},\,\ldots\,,\, x_{m}\right)\approx-\frac{1}{K}\ln\sum_{i=1}^{m}\texttt{e}^{-Kx_{i}}.\]
This smooth approximation converges to the minimum function as $K$
increases \cite{Vontobel-turbo06}. Since the soft-minimum function
is used in two different ways, we use different constants, $K_{1}$
and $K_{2},$ for the code and trellis terms. The smoothness of Problem-DS
allows one to to take derivative of \eqref{eq:JLP3-1} (giving the
Karush\textendash{}Kuhn\textendash{}Tucker (KKT) equations, derived
in Lemma \ref{lem:InnerLoop}), and represent \eqref{eq:JLP3-2} and
\eqref{eq:JLP3-3} using BCJR-like forward/backward recursions (given
by Lemma \ref{lem:OuterLoop}).

\begin{table}[b]
\captionsetup{name=Table} \caption{Dual Problem 2nd Formulation (Problem-D2)}\label{Flo:dual2}\[
\max_{\mathbf{m}}\sum_{j\in\emph{J}}\min_{\mathcal{B}\in\mathcal{E}_{j}}\!\left[\sum_{i\in\mathcal{B}}m_{i,j}\right]\!+\min_{e\in\mathcal{T}_{p}}\!\left[\Gamma_{p,e}\!-\!\overrightarrow{n}_{p\!-\!1,s(e)}\!+\!\overleftarrow{n}_{p,s'(e)}\right]\]
where $\overrightarrow{n}_{i,k}$ is defined for $i=1,\,\ldots\,,\, p-1$
by \vspace{0mm} \[
-\overrightarrow{n}_{i,k}=\min_{e\in s'^{-1}(k)}-\overrightarrow{n}_{i-1,s(e_{i})}+\Gamma_{i,e},\,\forall k\in\mathcal{S}\vspace{-1mm}\]
and $\overleftarrow{n}_{i,k}$ is defined for $i=N-1,N-2,\,\ldots\,,\, p$
by \vspace{0mm}\emph{\[
\overleftarrow{n}_{i,k}=\min_{e\in s^{-1}(k)}\overleftarrow{n}_{i+1,s'(e_{i+1})}+\Gamma_{i+1,e},\,\forall k\in\mathcal{S}\vspace{-1.5mm}\]
}starting from \vspace{-1mm}\[
\overrightarrow{n}_{0,k}=\overleftarrow{n}_{N,k}=0,\,\forall k\in\mathcal{S}.\]

\end{table}

\begin{table}[b]
\captionsetup{name=Table} \caption{Softened Dual Problem (Problem-DS)}\label{Flo:softdual}\begin{eqnarray}
\max_{\mathbf{m}}\,\,-\frac{1}{K_{1}}\sum_{j\in\emph{J}}\ln\sum_{\mathcal{B}\in\mathcal{E}_{j}}\texttt{e}^{-K_{1}\left\{ \sum_{i\in\mathcal{N}(j)}m_{i,j}\mathbbm{1}_{\mathcal{B}}(i)\right\} }\label{eq:JLP3-1}\\
-\frac{1}{K_{2}}\ln\sum_{e\in\mathcal{T}_{p}}\texttt{e}^{-K_{2}\left\{ \Gamma_{p,e}-\overrightarrow{n}_{p-1,s(e)}+\overleftarrow{n}_{p,s'(e)}\right\} }\nonumber \end{eqnarray}
where $\mathbbm{1}_{\mathcal{B}}\left(i\right)$ is the indicator
function of the set $\mathcal{B},$ $\overrightarrow{n}_{i,k}$ is
defined for $i=1,\,\ldots\,,\, p-1$ by\emph{ }\begin{equation}
-\overrightarrow{n}_{i,k}=-\frac{1}{K_{2}}\ln\!\!\sum_{e_{i}\in s'^{-1}(k)}\!\!\texttt{e}^{-K_{2}\left\{ -\overrightarrow{n}_{i-1,s(e_{i})}+\Gamma_{i,e}\right\} },\label{eq:JLP3-2}\end{equation}
and $\overleftarrow{n}_{i,k}$ is defined for $i=N-1,N-2,\,\ldots\,,\, p$
by\begin{align}
\overleftarrow{n}_{i,k} & =-\frac{1}{K_{2}}\ln\!\!\!\sum_{e_{i+1}\in s^{-1}(k)}\!\!\!\texttt{e}^{-K_{2}\left\{ \overleftarrow{n}_{i+1,s'(e_{i+1})}+\Gamma_{i+1,e}\right\} }\label{eq:JLP3-3}\end{align}

\selectlanguage{english}%
\noindent \vspace{-3mm}\foreignlanguage{american}{starting from\[
\overrightarrow{n}_{0,k}=\overleftarrow{n}_{N,k}=0,\,\forall k\in\mathcal{S}.\]
}\selectlanguage{american}

\end{table}

\begin{figure}[t]
\begin{centering}
\includegraphics[width=0.5\columnwidth]{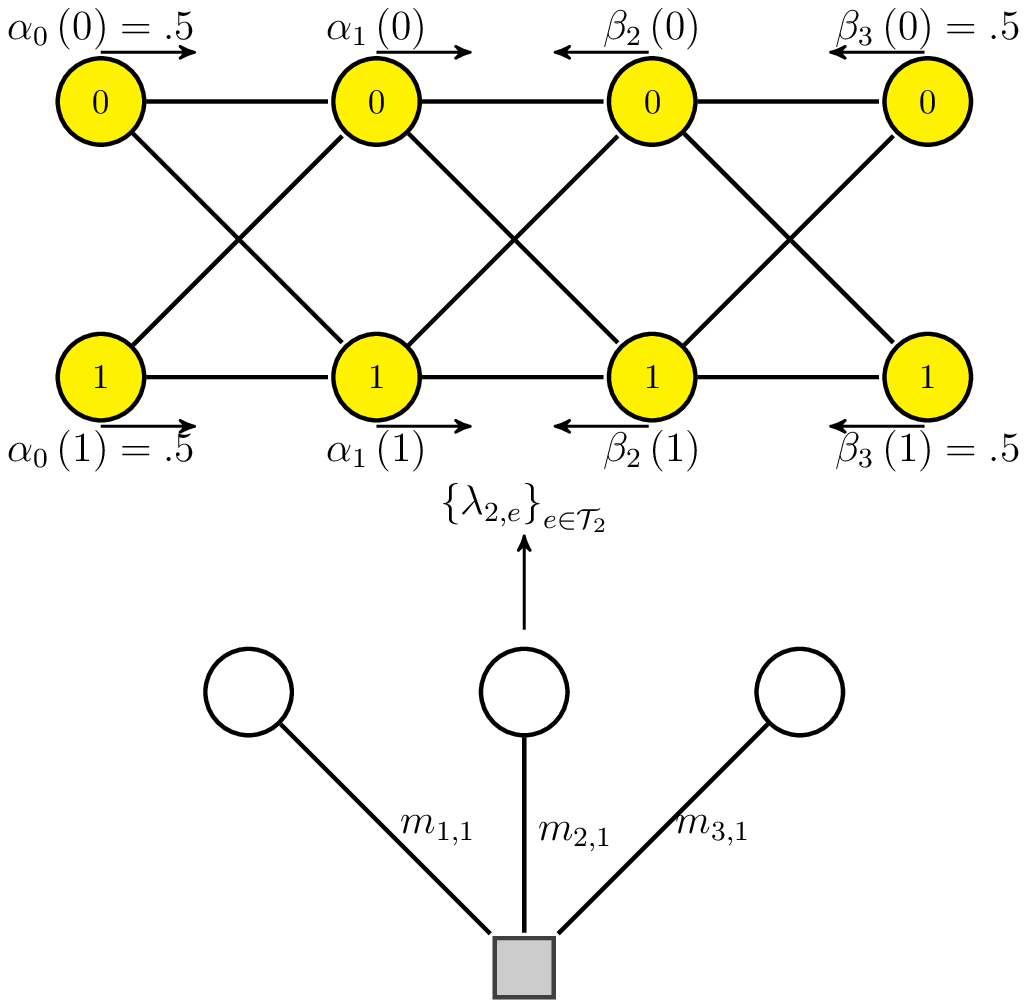}~\includegraphics[width=0.47\columnwidth]{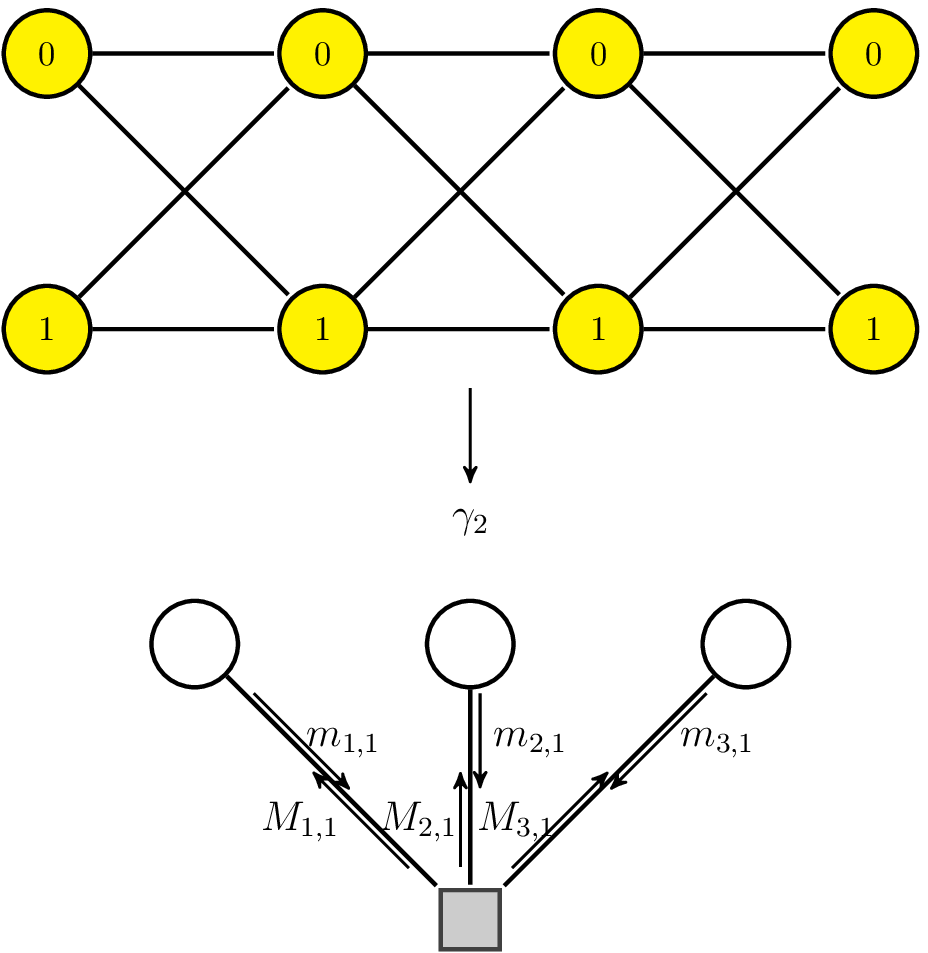}
\par\end{centering}

\caption{\label{fig:Loops} Illustration of Algorithm \ref{alg:IJLP} steps
for $i=2$ on the same example given by Fig. \ref{fig:Example}: outer
loop update (left) and inner loop update (right). }

\end{figure}

\begin{lem}
\emph{\label{lem:InnerLoop}Consider the KKT equations associated
with performing the minimization in \eqref{eq:JLP3-1} only over the
variables $\{m_{p,j'}\}_{j'\in\mathcal{N}(p)}$. These equations have
a unique solution given by}%
\begin{comment}
\emph{{[}The unique maximum of \eqref{eq:JLP3-1} over $\left\{ m_{p,j'}\right\} _{j'\in\mathcal{N}(p)}$
can be found using the KKT equations, and an iterative solution for}
\emph{$p\in\mathcal{I}$ is given by{]}}
\end{comment}
{}\emph{ }\textup{\[
m_{p,j'}=M_{p,j'}+\frac{\gamma_{p}}{K_{1}},\,\,\, M_{p,j'}\triangleq\frac{1}{K_{1}}\,\ln\,\frac{1-l_{p,j'}}{1+l_{p,j'}}\]
}\emph{for }$j'\in\mathcal{N}(p)$ \textup{where} \textup{\[
l_{p,j'}\triangleq\prod_{i\in\mathcal{N}(j')\setminus p}\tanh\left(\frac{K_{1}m_{i,j'}}{2}\right),\]
\[
\gamma_{p}\triangleq\ln\,\frac{\sum_{e\in\mathcal{T}_{p}:x(e)=0}\texttt{e}^{-K_{2}\left(\Gamma_{p}-\overrightarrow{n}_{p-1,s(e)}+\overleftarrow{n}_{p,s'(e)}\right)}}{\sum_{e\in\mathcal{T}_{p}:x(e)=1}\texttt{e}^{-K_{2}\left(\Gamma_{p}-\overrightarrow{n}_{p-1,s(e)}+\overleftarrow{n}_{p,s'(e)}\right)}}.\]
}\end{lem}
\begin{IEEEproof}
See Appendix \ref{sub:ProofInnerLoop}.\end{IEEEproof}
\begin{lem}
\label{lem:OuterLoop}\emph{ Equations \eqref{eq:JLP3-2} and \eqref{eq:JLP3-3}
are equivalent to the BCJR-based forward and backward recursion given
by (\ref{eq:BCJR1}), (\ref{eq:BCJR2}), and (\ref{eq:BCJR3}).}\end{lem}
\begin{IEEEproof}
By letting, $\alpha_{i}\left(k\right)\propto\texttt{e}^{K_{_{2}}\overrightarrow{n}_{i,k}},\,\lambda_{i+1,e}=\texttt{e}^{-K_{_{2}}\Gamma_{i+1,e}},\,\mbox{and}\,\beta_{i}\left(k\right)\propto\texttt{e}^{-K_{_{2}}\overleftarrow{n}_{i,k}}$,
we obtain the desired result by normalization.
\end{IEEEproof}
Now, we have all the pieces to complete the algorithm. As the last
step, we combine the results of Lemma \ref{lem:InnerLoop} and \ref{lem:OuterLoop}
to obtain the iterative solver for the joint-decoding LP, which is
summarized by the iterative joint LP decoding in Algorithm \ref{alg:IJLP}
(see Fig. \ref{fig:Loops} for a graphical depiction). %
{}
\begin{rem}
\label{rem:TEconnection}While Algorithm \ref{alg:IJLP} always has
a bit-node update rule different from standard belief propagation
(BP), we note that setting $K_{1}=1$ in the inner loop gives the
exact BP check-node update and setting $K_{2}=1$ in the outer loop
gives the exact BCJR channel update. In fact, one surprising result
of this work is that such a small change to the BCJR-based TE update
provides an iterative solver for the LP whose per-iteration complexity
similar to TE. It is also possible to prove the convergence of a slightly
modified iterative solver that is based on a less efficient update
schedule. %
\begin{algorithm}[t]
\caption{\label{alg:IJLP}Iterative Joint Linear-Programming Decoding}

\begin{itemize}
\item Step 1. Initialize $m_{i,j}=0$ for $i\in\mathcal{I},\, j\in\mathcal{N}\left(i\right)$
and $\ell=0.$
\item Step 2. Update Outer Loop: For $i\in\mathcal{I}$,

\begin{itemize}
\item (i) Compute bit-to-trellis message\vspace{-0.5mm} \[
\lambda_{i,e}=\texttt{e}^{-K_{_{2}}\Gamma_{i,e}}\vspace{-2mm}\]
where \[
\Gamma_{i,e}=b_{i,e}-\delta_{x(e)=1}\sum_{j\in\mathcal{N}(i)}m_{i,j}.\vspace{-1mm}\]
%
\begin{comment}
\[
\lambda_{i,e}=\texttt{e}^{-K_{_{2}}\left(b_{i,e}-\delta_{x(e)=1}\sum_{j\in\mathcal{N}(i)}m_{i,j}\right)}\]

\end{comment}
{}
\item (ii) Compute forward/backward trellis messages\vspace{-0.5mm}

\begin{equation}
\!\!\!\!\!\!\!\!\!\!\!\!\!\!\!\!\!\!\alpha_{i+1}\left(k\right)\!=\!\frac{\sum_{e\in s'^{-1}(k)}\alpha_{i}\left(s(e)\right)\cdot\lambda_{i+1,e}}{\sum_{k}\sum_{e\in s'^{-1}(k)}\alpha_{i}\left(s(e)\right)\cdot\lambda_{i+1,e}}\label{eq:BCJR1}\end{equation}
\begin{equation}
\!\beta_{i-1}\left(k\right)\!=\!\frac{\sum_{e\in s^{-1}(k)}\beta_{i}\left(s'(e)\right)\cdot\lambda_{i,e}}{\sum_{k}\sum_{e\in s^{-1}(k)}\beta_{i}\left(s'(e)\right)\cdot\lambda_{i,e}},\label{eq:BCJR2}\end{equation}
where $\beta_{N}\left(k\right)=\alpha_{0}\left(k\right)=1/\left|\mathcal{S}\right|$
for all $k\in\mathcal{S}$. 

\item (iii) Compute trellis-to-bit message $\gamma_{i}$\vspace{-1mm}\begin{equation}
\!\!\!\!\!\!\!\!\gamma_{i}\!=\!\mbox{\ensuremath{\ln}}\,\frac{\sum_{e\in\mathcal{T}_{i}:x(e)=0}\alpha_{i-1}\left(s(e)\right)\lambda_{i,e}\beta_{i}\left(s'(e)\right)}{\sum_{e\in\mathcal{T}_{i}:x(e)=1}\alpha_{i-1}\left(s(e)\right)\lambda_{i,e}\beta_{i}\left(s'(e)\right)}\label{eq:BCJR3}\end{equation}

\end{itemize}
\item Step 3. Update Inner Loop for $\ell_{\mbox{inner}}$ rounds: For $i\in\mathcal{I}$,

\begin{itemize}
\item (i) Compute bit-to-check msg $m_{i,j}$ for $j\in\mathcal{N}\left(i\right)$\vspace{0.5mm}
\[
m_{i,j}=M_{i,j}+\frac{\gamma_{i}}{K_{1}}\]

\item (ii) Compute check-to-bit msg $M_{i,j}$ for $j\in\mathcal{N}\left(i\right)$\vspace{0.5mm}
\begin{equation}
M_{i,j}=\frac{1}{K_{1}}\,\ln\,\frac{1-l_{i,j}}{1+l_{i,j}}\label{eq:chk1}\end{equation}
where\vspace{0mm} \begin{equation}
l_{i,j}=\prod_{r\in\mathcal{N}\left(j\right)\setminus i}\tanh\left(\frac{K_{1}m_{r,j}}{2}\right)\label{eq:chk2}\end{equation}
%
\begin{comment}
\[
M_{i,j}=-\frac{2}{K_{1}}\mbox{atanh}\left(\prod_{r\in\mathcal{N}\left(j\right)\setminus i}\tanh\left(\frac{K_{1}m_{r,j}}{2}\right)\right)=-\frac{2}{K_{1}}\mbox{tanh}^{-1}\left(\prod_{r\in\mathcal{N}\left(j\right)\setminus i}\tanh\left(\frac{K_{1}m_{r,j}}{2}\right)\right)\]
$|N(j)|$
\end{comment}
{}
\end{itemize}
\item Step 4. Compute hard decisions and stopping rule \vspace{0mm}

\begin{itemize}
\item (i) For $i\in\mathcal{I}$,\vspace{0mm} \begin{align*}
\hat{f}_{i} & =\begin{cases}
1 & \mbox{if}\,\,\,\gamma_{i}<0\\
0, & \mbox{otherwise}\end{cases}\end{align*}

\item (ii) If $\mathbf{\hat{f}}$ satisfies all parity checks or the maximum
outer iteration number, $\ell_{\mbox{outer}}$, is reached, stop and
output $\mathbf{\hat{f}}$. Otherwise increment $\ell$ and go to
Step 2.
\end{itemize}
\end{itemize}

\end{algorithm}

\end{rem}

\subsection{Convergence Analysis \label{sub:Convergence}}

%
\begin{comment}
This figure shows comparison between the joint LP decoding (JLPD),
joint iterative message-passing decoding (JIMPD), and iterative joint
LP decoding (IJLPD) on the pDIC with AWGN for random (3,5) regular
LDPC codes of length $N=155$ (left) and $N=450$ (right). The curves
shown are the JLPD WER (solid), JLPD WER prediction (dashed), JIMPD
WER (dash-dot), and IJLPD WER (circle-solid). The JLPD experiments
were repeated for three different non-zero codewords in each case.
The dashed curves are computed using the union bound in Equation \eqref{eq:UnionBound}
based on JD-PCWs observed at 3.46~dB (left) 2.67~dB (right) and
the dash-dot curves are obtained using the state-based JIMPD described
in \cite{Kavcic-it03}. The circle-solid curves are computed using
Algorithm \ref{alg:IJLP}. Note that SNR is defined as channel output
power divided by $\sigma^{2}$.
\end{comment}
{}This section considers the convergence properties of Algorithm \ref{alg:IJLP}.
Although simulations have not shown any convergence problems with
Algorithm \ref{alg:IJLP} in its current form, our proof requires
a modified update schedule that is less computationally efficient.
%
\begin{comment}
This convergence properties will remove the pain to adjust number
of inner and outer loops for running Algorithm \ref{alg:IJLP}, unlike
the standard TE approaches as discussed in Section \ref{sec:Simulations}.
\end{comment}
{}Following Vontobel's approach in \cite{Vontobel-turbo06}, which is
based on general properties of Gauss-Seidel-type algorithms for convex
minimization, we show that the modified version Algorithm \ref{alg:IJLP}
is guaranteed to converge. Moreover, a feasible %
\begin{comment}
primal 
\end{comment}
{}solution to Problem-P can be obtained whose value is arbitrarily close
to the optimal value of Problem-P. %
\begin{comment}
LP solution of Problem with a computational complexity which scales
linearly with the block length%
\footnote{For proof purpose, we take simpler scheduling approach, extending
them to different schedules will be tedious but can be done. %
}. 
\end{comment}
{}

The modified update rule for Algorithm \ref{alg:IJLP} consists of
cyclically, for each $p=1,\ldots,N$, computing the quantity $\gamma_{p}$
(via step 2 of Algorithm \ref{alg:IJLP}) and then updating $m_{p,j}$
for all $j\in\mathcal{N}(p)$ (based on step 3 of Algorithm \ref{alg:IJLP}).
The drawback of this approach is that one BCJR update is required
for each bit update, rather than for $N$ bit updates. This modification
allows us to interpret Algorithm \ref{alg:IJLP} as a Gauss-Seidel-type
algorithm. We believe that, at the expense of a longer argument, the
convergence proof can be extended to a decoder which uses windowed
BCJR updates (e.g., see \cite{Kavcic-it03}) to achieve convergence
guarantees with much lower complexity. Regardless, the next few lemmas
and theorems can be seen as a natural generalization of \cite{Vontobel-turbo06}\cite{Burshtein-it09}
to the joint-decoding problem.

\begin{table}[b]
\captionsetup{name=Table} \caption{Softened Primal Problem (Problem-PS)}\label{Flo:softprimal}\[
\min_{\mathbf{g,w}}\sum_{i\in\mathcal{I}}\sum_{e\in\mathcal{T}_{i}}b_{i,e}g_{i,e}-\frac{1}{K_{1}}\sum_{j\in\mathcal{J}}H(w_{j})-\frac{1}{K_{2}}H(g_{p})\]

\selectlanguage{english}%
\vspace{0mm}

\selectlanguage{american}%
subject to the same constraints as Problem-P.
\end{table}

\begin{lem}
\emph{\label{lem:convergence}Assume that all the rows of $H$ have
Hamming weight at least 3. Then, the modified Algorithm \ref{alg:IJLP}
converges to the maximum of the Problem-DS.}\end{lem}
\begin{IEEEproof}
See Appendix \ref{sub:Proofconvergence}.
\end{IEEEproof}
Next, we introduce the softened primal problem (Problem-PS) in Table
\ref{Flo:softprimal}, using the definitions $w_{j}\triangleq\left\{ w_{j,\mathcal{B}}\right\} _{\mathcal{B}\in\mathcal{E}_{j}}$
and $g_{p}\triangleq\left\{ g_{p,e}\right\} _{e\in\mathcal{T}_{p}}$.
Using standard convex analysis (e.g., see \cite[p. 254, Ex. 5.5]{Boyd-2004}),
one can show that Problem-PS is the Lagrangian dual of Problem-DS
and that the minimum of Problem-PS is equal to the maximum of Problem-DS.\emph{
}In particular, Problem-PS can be seen as a maximum-entropy regularization
of Problem-DS that was derived by smoothing dual problem given by
Problem-D2. Thus, our Algorithm \ref{alg:IJLP} is dually-related
to an interior-point method for solving the LP relaxation of joint
ML decoding on trellis-wise polytope using the entropy function (for
$x$ in the standard simplex)\begin{equation}
H(x)\triangleq-\sum_{i}x_{i}\,\ln x_{i}\label{eq:entropy_def}\end{equation}
as a barrier function (e.g., see \cite[p. 126]{Johnson-2008}) for
the polytope. 
\begin{rem}
\label{rem:free-energy}By taking sufficiently large $K_{1}$ and
$K_{2}$\emph{, }the primal LP of joint LP decoder\emph{ }in Problem-P,
emerges as the {}``zero temperature'' limit of the approximate LP
relaxations given by Problem-PS \cite{Vontobel-turbo06}\cite{Johnson-2008}.
Also, Problem-PS can be seen as a convex free-energy minimization
problem \cite{Johnson-2008}.
\end{rem}
Next, we develop a relaxation bound\emph{, }given by Lemma \ref{lem:relaxation bound}
and Lemma \ref{lem:Fesability} to quantify the performance loss of
Algorithm \ref{alg:IJLP} (when it converges) in relation to the joint
LP decoder.
\begin{lem}
\emph{\label{lem:relaxation bound}Let $P^{*}$ be the minimum value
of Problem-P and $\tilde{P}$ be the minimum value of Problem-PS.}%
\begin{comment}
{[}gap condition: \emph{$P\left(\mathbf{g}^{K}\right)-P\left(\mathbf{p}^{*}\right)>0?]$}
\end{comment}
{}\emph{ Then \[
0\leq\tilde{P}-P^{*}\leq\delta N,\]
where \[
\bar{\mathcal{N}}\triangleq\frac{\sum_{j\in\mathcal{J}}\left|\mathcal{N}(j)\right|}{N},\, R\triangleq1-\frac{M}{N}\]
and \[
\delta\triangleq\frac{\left(1-R+\bar{\mathcal{N}}\right)\ln\,2}{K_{1}}+\frac{\ln\, O}{K_{2}N}.\]
}\end{lem}
\begin{IEEEproof}
See Appendix \ref{sub:Proofrelaxationbnd}.\end{IEEEproof}
\begin{lem}
\emph{\label{lem:Fesability}For any $\epsilon>0$},\emph{ the modified
Algorithm \ref{alg:IJLP} returns a feasible solution for Problem-DS
that satisfies the KKT conditions within $\epsilon$. With this, one
can construct a feasible solution $\left(\mathbf{\tilde{g}}_{\epsilon},\mathbf{\tilde{w}}_{\epsilon}\right)$
for Problem-PS that has the (nearly optimal) value $\tilde{P}_{\epsilon}$.
For small enough $\epsilon$, one finds that }\[
0\leq\tilde{P}_{\epsilon}-\tilde{P}\leq\delta N,\]
\emph{where} \[
\delta\triangleq\frac{\left(1-R+\bar{\mathcal{N}}\right)\ln\,2}{K_{1}}+\epsilon\left(\!\frac{3}{N}\!\sum_{l\in\mathcal{I}}\!\sum_{\substack{e\in\mathcal{T}_{l}}
}\left|b_{l,e}\right|+C\right).\]
\end{lem}
\begin{IEEEproof}
See Appendix \ref{sub:Prooffeasibility}.
\end{IEEEproof}
Lastly, we obtain the desired conclusion, which is stated as Theorem
\ref{thm:Fesability}.
\begin{thm}
\emph{\label{thm:Fesability}For any $\delta>0$},\emph{ there exists
a sufficiently small $\epsilon>0$ and sufficiently large $K_{1}$
and $K_{2}$ such that finitely many iterations of the modified Algorithm
\ref{alg:IJLP} can be used to construct a feasible $\left(\mathbf{\tilde{g}}_{\epsilon},\mathbf{\tilde{w}}_{\epsilon}\right)$
for Problem-PS that is also nearly optimal. The value of this solution
is denoted $\tilde{P}_{\epsilon}$ and satisfies }\[
0\leq\tilde{P}_{\epsilon}-P^{*}\leq\delta N,\]
\emph{where} \[
\delta\triangleq\frac{\left(1-R+\bar{\mathcal{N}}\right)\ln\,2}{K_{1}}+\frac{\ln\, O}{K_{2}N}+\epsilon\left(\!\frac{3}{N}\!\sum_{l\in\mathcal{I}}\!\sum_{\substack{e\in\mathcal{T}_{l}}
}\left|b_{l,e}\right|+C\right).\]
\end{thm}
\begin{IEEEproof}
Combining results of Lemma \ref{lem:convergence}, Lemma \ref{lem:relaxation bound},
and Lemma \ref{lem:Fesability}, we obtain the desired error bound%
\begin{comment}
 of $\delta$
\end{comment}
{}. \end{IEEEproof}
\begin{rem}
The modified (i.e., cyclic schedule) Algorithm \ref{alg:IJLP} is
guaranteed to converge to a solution whose value can be made arbitrarily
close to $P^{*}.$ %
\begin{comment}
The distance, normalized by the block length, between the minimum
value and the value of the solution provided by Algorithm \ref{alg:IJLP}
can be made arbitrarily small.
\end{comment}
{} Therefore, the joint iterative LP decoder provides an approximate
solution to Problem-P whose value is governed by the upper bound in
Theorem \ref{thm:Fesability}. %
\begin{comment}
However, numerical precision limits the maximum values of \emph{$K_{1}$
}and \emph{$K_{2},$} which limit the practical size of $\delta$.
\end{comment}
{} Algorithm \ref{alg:IJLP} can be further modified to be of Gauss-Southwell
type so that the complexity analysis in \cite{Burshtein-it09} can
be extended to this case. Still, the analysis in \cite{Burshtein-it09},
although a valid upper bound, does not capture the true complexity
of decoding because one must choose $\delta=o\left(\frac{1}{N}\right)$
to guarantee that the iterative LP solver finds the true minimum.
Therefore, the exact convergence rate and complexity analysis of Algorithm
\ref{alg:IJLP} is left for future study. In general, the convergence
rate of coordinate-descent methods (e.g., Gauss-Seidel and Gauss-Southwell
type algorithms) for convex problems without strict convexity is an
open problem. %
{}
\end{rem}

\section{Error Rate Prediction and Validation\label{sec:Simulations} }

In this section, we validate the proposed joint-decoding solution
and discuss some implementation issues. Then, we present simulation
results and compare with other approaches. In particular, we compare
the performance of the joint LP decoder and joint iterative LP decoder
with the joint iterative message-passing decoder on two finite-state
intersymbol interference channels (FSISCs) described in Definition
\ref{def:exFSISI}. For preliminary studies, we use a $(3,\,5)$-regular
binary LDPC code on the precoded dicode channel (pDIC) with length
155 and 455. For a more practical scenario, we also consider a $(3,\,27)$-regular
binary LDPC code with length 4923 and rate 8/9 on the class-II Partial
Response (PR2) channel used as a partial-response target for perpendicular
magnetic recording. All parity-check matrices were chosen randomly
except that double-edges and four-cycles were avoided. Since the performance
depends on the transmitted codeword, the WER results were obtained
for a few chosen codewords of fixed weight. The weight was chosen
to be roughly half the block length, giving weights 74, 226, and 2462
respectively.

The performance of the three algorithms was assessed based on the
following implementation details.

\paragraph*{Joint LP Decoder}

Joint LP decoding is performed in the dual domain because this is
much faster than the primal domain when using MATLAB. Due to the slow
speed of LP solver, simulations were completed up to a WER of roughly
$10^{-4}$ on the three different non-zero LDPC codes with block lengths
155 and 455 each. To extrapolate the error rates to high SNR (well
beyond the limits of our simulation), we use a simulation-based semi-analytic
method with a truncated union bound (see \eqref{eq:UnionBound}) as
discussed in Section \ref{sec:Joint-LPD}. The idea is to run a simulation
at low SNR and keep track of all observed codeword and pseudo-codeword
(PCW) errors and a truncated union bound is computed by summing over
all observed errors. The truncated union bound is obtained by computing
the generalized Euclidean distances associated with all decoding errors
that occurred at some low SNR points (e.g., WER of roughly than $10^{-1}$)
until we observe a stationary generalized Euclidean distance spectrum.
\foreignlanguage{english}{It is quite easy, in fact, to store these
error events in a list which is finally pruned to avoid overcounting.
Of course, low SNR allows the decoder to discover PCWs more rapidly
than high SNR }and it is well-known that the truncated bound should
give a good estimate at high SNR if all dominant joint decoding PCWs
have been found (e.g., \cite{Hu-itm10,Lee-isit08}).\foreignlanguage{english}{
One nontrivial open question is the feasibility and effectiveness
of enumerating error events for long codes. In particular, we do not
address how many instances must be simulated to have high confidence
that all the important error events are found so there are no surprises
at high SNR.}%
\begin{figure*}[t]
\begin{centering}
\includegraphics[width=0.27\paperheight]{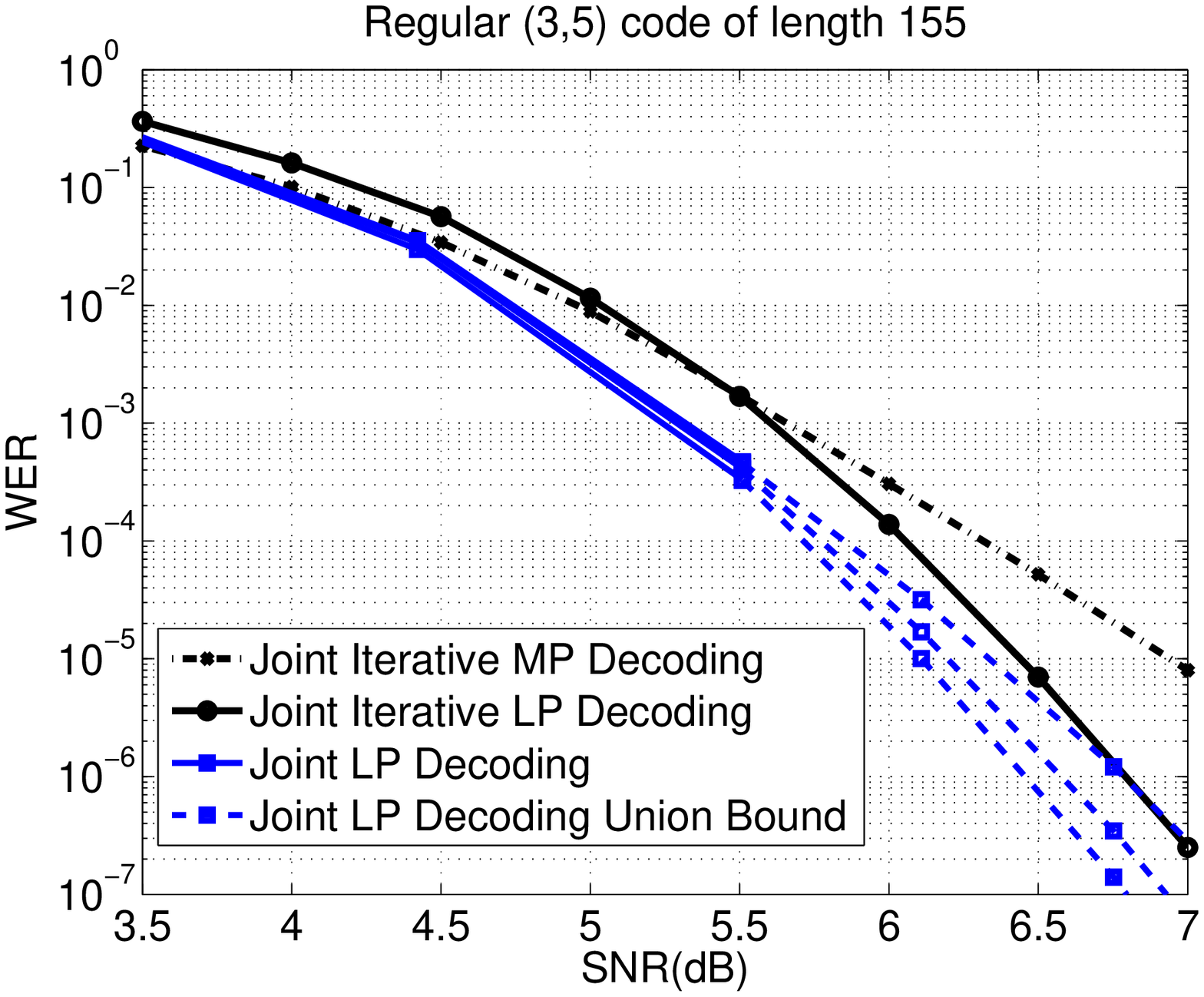}~~~\includegraphics[width=0.27\paperheight]{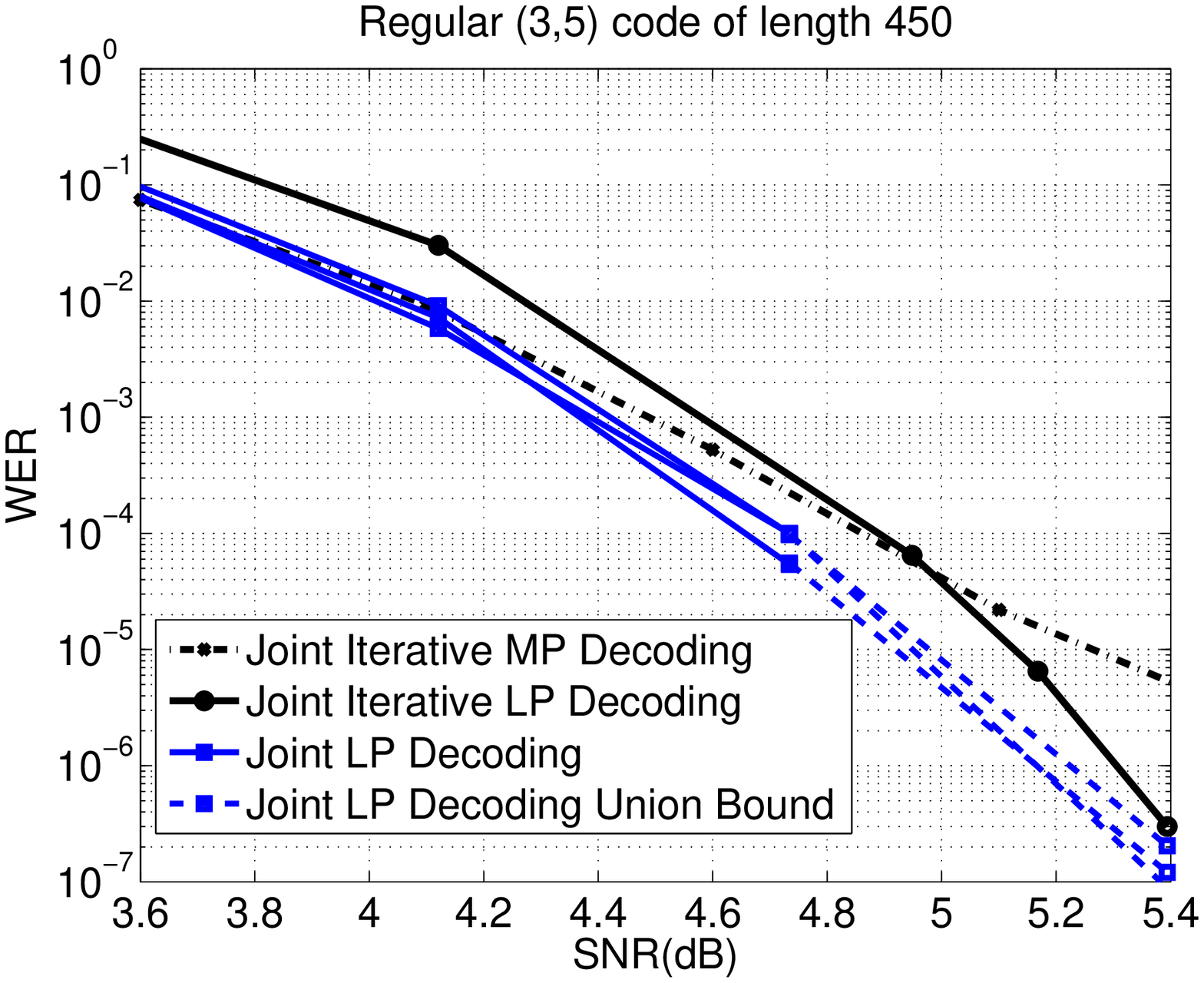}
\par\end{centering}

\caption{\label{fig:result1}Comparisons between the joint LP decoding, joint
iterative LP decoding, and joint iterative message-passing (MP) decoding
on the pDIC with AWGN for random (3,5) regular LDPC codes of length
$N=155$ (left) and $N=450$ (right). The joint LP decoding experiments
were repeated for three different non-zero codewords and depicted
in three different curves. The dashed curves are computed using the
union bound in Equation \eqref{eq:UnionBound} based on JD-PCWs observed
at 3.46~dB (left) 2.67~dB (right). Note that SNR is defined as channel
output power divided by $\sigma^{2}$.}

\end{figure*}

\paragraph*{Joint Iterative LP Decoder}

Joint iterative decoding is performed based on the Algorithm \ref{alg:IJLP}
on all three LDPC codes of different lengths. For block lengths 155
and 455, we chose the codeword which shows the worst performance for
the joint LP decoder experiments. We used a simple scheduling update
scheme: variables are updated according to Algorithm \ref{alg:IJLP}
with cyclically with $\ell_{\mbox{inner}}=2$ inner loop iterations
for each outer iteration. The maximum number of outer iterations is
$\ell_{\mbox{outer}}=100$, so the total iteration count, $\ell_{\mbox{outer}}\ell_{\mbox{inner}}$,
is at most 200. The choice of parameters are $K_{1}=1000$ and \foreignlanguage{english}{$K_{2}=100$}
on the LDPC codes with block lengths 155 and 455. For the LDPC code
with length 4923, $K_{2}$ is reduced to 10. To prevent possible underflow
or overflow, a few expressions must be implemented carefully. When
\[
K_{1}\min_{r\in\mathcal{N}\left(j\right)\setminus i}m_{r,j}\geq35,\]
a well-behaved approximation of \eqref{eq:chk1} and \eqref{eq:chk2}
is given by \begin{multline*}
\!\!\!\!\!\!\!\!\left[\!\frac{1}{K_{1}}\!\ln\!\left\{ \!\left(\!2\!+2\!\!\!\!\!\sum_{r\in\mathcal{N}\left(j\right)\setminus i}\!\texttt{e}^{-K_{1}\left(|m_{r,j}|-\min_{r\in\mathcal{N}\left(j\right)\setminus i}m_{r,j}\right)}\right)\right\} \right.\\
\left.-\min_{r\in\mathcal{N}\left(j\right)\setminus i}m_{r,j}\right]\mbox{sgn}\left(l_{i,j}\right),\end{multline*}
where $\mbox{sgn}\left(x\right)$ is the usual sign function. Also,
\eqref{eq:BCJR3} should be implemented using\begin{multline*}
\max_{e\in\mathcal{T}_{i}:x(e)=0}\left\{ \bar{\alpha}_{i-1}\left(s(e)\right)+\bar{\lambda}_{i,e}+\bar{\beta}_{i}\left(s'(e)\right)\right\} \\
-\max_{e\in\mathcal{T}_{i}:x(e)=1}\left\{ \bar{\alpha}_{i-1}\left(s(e)\right)+\bar{\lambda}_{i,e}+\bar{\beta}_{i}\left(s'(e)\right)\right\} \\
+\mbox{log}\,\left[\sum_{e\in\mathcal{T}_{i}:x(e)=0}\bar{\alpha}_{i-1}\left(s(e)\right)+\bar{\lambda}_{i,e}+\bar{\beta}_{i}\left(s'(e)\right)-\right.\\
\left.\max_{e\in\mathcal{T}_{i}:x(e)=0}\left\{ \bar{\alpha}_{i-1}\left(s(e)\right)+\bar{\lambda}_{i,e}+\bar{\beta}_{i}\left(s'(e)\right)\right\} \right]\\
-\mbox{log}\,\left[\sum_{e\in\mathcal{T}_{i}:x(e)=1}\bar{\alpha}_{i-1}\left(s(e)\right)+\bar{\lambda}_{i,e}+\bar{\beta}_{i}\left(s'(e)\right)-\right.\\
\left.\max_{e\in\mathcal{T}_{i}:x(e)=1}\left\{ \bar{\alpha}_{i-1}\left(s(e)\right)+\bar{\lambda}_{i,e}+\bar{\beta}_{i}\left(s'(e)\right)\right\} \right],\end{multline*}
where $\bar{\alpha}_{i}\left(k\right)\triangleq\ln\alpha_{i}\left(k\right),\,\bar{\beta}_{i}\left(k\right)\triangleq\ln\beta_{i}\left(k\right)$
and $\bar{\lambda}_{i,e}\triangleq\ln\lambda_{i,e}.$

\paragraph*{Joint Iterative Message-Passing Decoder}

Joint iterative message decoding is performed based on the state-based
algorithm described in \cite{Kavcic-it03} on all three LDPC codes
of different lengths. To make a fair comparison with the Joint Iterative
LP Decoder, the same maximum iteration count and the same codewords
are used.

\subsection{Results}

Fig. \ref{fig:result1} compares the results of all three decoders
and the error-rate estimate given by the union bound method discussed
in Section \ref{sec:Joint-LPD}. The solid lines represent the simulation
curves while the dashed lines represent a truncated union bound for
three different non-zero codewords. Surprisingly, we find that joint
LP decoder outperforms joint iterative message passing decoder by
about 0.5~dB at WER of $10^{-4}$. We also observe that that joint
iterative LP decoder loses about 0.1~dB at low SNR. This may be caused
by using finite values for $K_{1}$ and $K_{2}$. At high SNR, however,
this gap disappears and the curve converges towards the error rate
predicted for joint LP decoding. This shows that joint LP decoding
outperforms belief-propagation decoding for short length code at moderate
SNR with the predictability of LP decoding. Of course, this can be
achieved with a computational complexity similar to turbo equalization. 

One complication that must be discussed is the dependence on the transmitted
codeword. Computing the bound is complicated by the fact that the
loss of channel symmetry implies that the dominant PCWs may depend
on the transmitted sequence. It is known that long LDPC codes with
joint iterative decoding experience a concentration phenomenon \cite{Kavcic-it03}
whereby the error probability of a randomly chosen codeword is very
close, with high probability, to the average error probability over
all codewords. This effect starts to appear even at the short block
lengths used in this example. More research is required to understand
this effect at moderate block lengths and to verify the same effect
for joint LP decoding. 

\begin{figure}[t]
\begin{centering}
\includegraphics[width=0.27\paperheight]{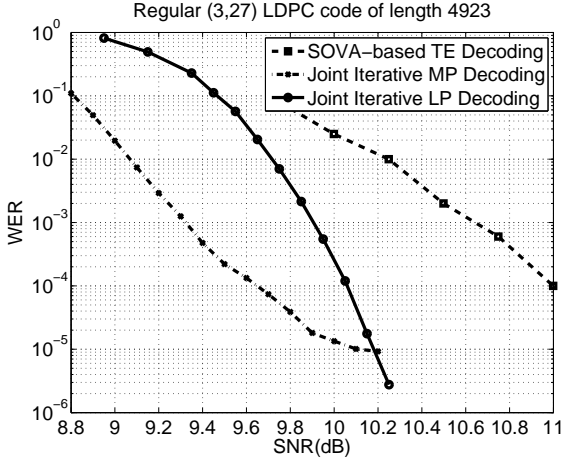}
\par\end{centering}

\caption{\label{fig:result3}Comparisons between the joint iterative LP decoding,
joint iterative MP decoding and soft-output Viterbi algorithm (SOVA)-based
TE decoding (taken from \cite{Jeon-itm07}) on the PR2 channel with
AWGN for random (3,27) regular LDPC codes of length $N=4923$. Note
that SNR is defined as channel output power divided by $\sigma^{2}$.}

\end{figure}

Fig. \ref{fig:result3} compares the joint iterative LP decoder and
joint iterative message-passing decoder in a practical scenario. Again,
we find that the joint iterative LP decoder provides gains over the
joint iterative message-passing decoder at high SNR. The slope difference
between the curves also suggests that the performance gains of joint
iterative LP decoder will increase with SNR. This shows that joint
iterative LP decoding can provide performance gains at high SNR with
a computational complexity similar to that of turbo equalization.

\section{Conclusions \label{sec:Concl}}

In this paper, we consider the problem of linear-programming (LP)
decoding of low-density parity-check (LDPC) codes and finite-state
channels (FSCs). First, we present an LP formulation of joint-decoding
for LDPC codes on FSCs that offers decoding performance improvements
over joint iterative message-passing decoding at moderate SNR. Then,
joint-decoding pseudo-codewords (JD-PCWs) are defined and the decoder
error rate is upper bounded by a union bound over JD-PCWs that is
evaluated for deterministic ISI channels with AWGN. Next, we propose
a simulation-based semi-analytic method for estimating the error rate
of LDPC codes on finite-state intersymbol interference channel (FSISIC)
at high SNR using only simulations at low SNR. Finally, we present
a novel iterative solver for the joint LP decoding problem. This greatly
reduces the computational complexity of the joint LP solver by exploiting
the LP dual problem structure. Its main advantage is that it provides
the predictability of LP decoding and significant gains over turbo
equalization (TE) especially in the error-floor with a computational
complexity similar to TE.

\appendix

\section{Proof Details}

\subsection{Proof of Lemma \ref{lem:InnerLoop}\label{sub:ProofInnerLoop}}

%
\begin{comment}
Consider the minimization in \eqref{eq:JLP3-1} over just the $p$-th
coordinate, $\left\{ m_{p,j}\right\} _{j\in\mathcal{N}(p)}$. This
gives for\emph{ $p\in\mathcal{I}$, \eqref{eq:JLP3-1} }is equivalent
to
\end{comment}
{} Restricting the minimization in \eqref{eq:JLP3-1} to the variables
$\{m_{p,j'}\}_{j'\in\mathcal{N}(p)}$ gives \begin{align}
 & -\!\!\!\min_{\left\{ m_{p,j}\right\} _{j\in\mathcal{N}(p)}}\left\{ \frac{1}{K_{1}}\sum_{j\in\mathcal{N}(p)}\ln\sum_{\mathcal{B}\in\mathcal{E}_{j}}\texttt{e}^{-K_{1}\sum_{i\in\mathcal{N}(j)}m_{i,j}\mathbbm{1}_{\mathcal{B}}(i)}+\right.\nonumber \\
 & \,\,\,\left.\frac{1}{K_{2}}\ln\sum_{e\in\mathcal{T}_{p}}\texttt{e}^{-K_{2}\left(\Gamma_{p,e}-\overrightarrow{n}_{p-1,s(e)}+\overleftarrow{n}_{p,s'(e)}\right)}\right\} .\label{eq:JLP4-1}\end{align}
The solution to \eqref{eq:JLP4-1} can be obtained by solving the
KKT equations. For $p\in\mathcal{I}$, we take the first derivative
with respect to $\left\{ m_{p,j'}\right\} _{j'\in\mathcal{N}(p)}$
and set it to zero; this yields

\begin{align}
 & \left(\frac{\sum_{\mathcal{B}\in\mathcal{E}_{j'},p\notin\mathcal{B}}\texttt{e}^{-K_{1}\sum_{i\in\mathcal{N}(j')\setminus p}m_{i,j'}\mathbbm{1}_{\mathcal{B}}(i)}}{\sum_{\mathcal{B}\in\mathcal{E}_{j'},\emph{B}\ni p}\texttt{e}^{-K_{1}\sum_{i\in\mathcal{N}(j')\setminus p}m_{i,j'}\mathbbm{1}_{\mathcal{B}}(i)}}\right)\cdot\texttt{e}^{K_{1}m_{p,j'}}=\nonumber \\
 & \left(\frac{\sum_{e\in\mathcal{T}_{p}:x(e)=0}\texttt{e}^{-K_{2}\left(\Gamma_{p,e}-\overrightarrow{n}_{p-1,s(e)}+\overleftarrow{n}_{p,s'(e)}\right)}}{\sum_{e\in\mathcal{T}_{p}:x(e)=1}\texttt{e}^{-K_{2}\left(\Gamma_{p,e}-\overrightarrow{n}_{p-1,s(e)}+\overleftarrow{n}_{p,s'(e)}\right)}}\right)\label{eq:Update2}\end{align}
By defining $-K_{1}M_{p,j'}$ as \begin{align}
 & \ln\frac{\sum_{\mathcal{B}\in\mathcal{E}_{j'},p\notin\mathcal{B}}\texttt{e}^{-K_{1}\sum_{i\in\mathcal{N}(j')\setminus p}m_{i,j'}\mathbbm{1}_{\mathcal{B}}(i)}}{\sum_{\mathcal{B}\in\mathcal{E}_{j'},\mathcal{B}\ni p}\texttt{e}^{-K_{1}\sum_{i\in\mathcal{N}(j')\setminus p}m_{i,j'}\mathbbm{1}_{\mathcal{B}}(i)}}\label{eq:definition M}\\
= & \ln\frac{\prod_{i\in\mathcal{N}(j')\setminus p}\left(1+\nu_{i,j'}\right)+\prod_{i\in\mathcal{N}(j')\setminus p}\left(1-\nu_{i,j'}\right)}{\prod_{i\in\mathcal{N}(j')\setminus p}\left(1+\nu_{i,j'}\right)-\prod_{i\in\mathcal{N}(j')\setminus p}\left(1-\nu_{i,j'}\right)}\nonumber \\
= & -\ln\frac{1-l_{p,j'}}{1+l_{p,j'}},\nonumber \end{align}
where $\nu_{i,j'}\triangleq\texttt{e}^{-K_{1}m_{i,j'}},$ we can rewrite
\eqref{eq:Update2} to obtain the desired result.

\subsection{Proof of Lemma \ref{lem:convergence}\label{sub:Proofconvergence}}

To characterize the convergence of the iterative joint LP decoder,
we consider the modification of Algorithm \ref{alg:IJLP} with cyclic
updates. The analysis follows \cite{Vontobel-turbo06} and uses the
proposition about \emph{convergence of} \emph{block coordinate descent
methods} from \cite[p. 247]{Bertsekas-1995}.
\begin{prop}
\label{pro:convergence}\emph{Consider the problem \[
\min_{x\in\mathcal{X}}\, f\left(x\right)\]
where $\mathcal{X}=\mathcal{X}_{1}\times\mathcal{X}_{2}\times\,\cdots\,\times\mathcal{X}_{m}$
and each $\mathcal{X}_{i}$ is a closed convex subset of $\mathbb{R}^{n_{i}}$.
The vector $x$ is partitioned so $x=\left(x_{1},\, x_{2},\,\ldots\,,\, x_{m}\right)$
with $x_{i}\in\mathbb{R}^{n_{i}}$. Suppose that $f$ is }continuously
differentiable and convex\emph{ on $\mathcal{X}$ and that, for every
$x\in\mathcal{X}$ and every $i=1,\ldots,m$, the problem \[
\min_{\xi_{i}\in\mathcal{X}_{i}}\, f\left(x_{1},\,\ldots\,,\, x_{i-1},\,\xi_{i},\, x_{i+1},\,\ldots\,,\, x_{m}\right)\]
has a }unique minimum\emph{. Now, consider the sequence $x^{k+1}=\left(x_{1}^{k+1},\,\ldots\,,\, x_{m}^{k+1}\right)$
defined by\[
x_{i}^{k+1}=\argmin_{\xi_{i}\in\mathcal{X}_{i}}f\left(x_{1}^{k+1},\,\ldots\,,\, x_{i-1}^{k+1},\,\xi_{i},\, x_{i+1}^{k},\,\ldots\,,\, x_{m}^{k}\right),\]
for $i=1,\ldots,m$. Then, every limit point of this sequence minimizes
$f$ over $\mathcal{X}.$}
\end{prop}
By using Proposition \ref{pro:convergence}, we will show that the
modified Algorithm \ref{alg:IJLP}\emph{ }converges. Define $\mathbf{m}_{i}=\left\{ m_{i,j}\right\} _{j\in\mathcal{N}(i)}$
and \begin{align*}
f\left(\mathbf{m}\right)\triangleq & f\left(\mathbf{m}_{1},\,\ldots\,,\,\mathbf{m}_{N}\right)\\
= & \frac{1}{K_{1}}\sum_{j\in\emph{J}}\ln\sum_{\mathcal{B}\in\mathcal{E}_{j}}\texttt{e}^{-K_{1}\left\{ \sum_{i\in\mathcal{N}(j)}m_{i,j}\mathbbm{1}_{\mathcal{B}}(i)\right\} }+\\
 & \frac{1}{K_{2}}\ln\sum_{e\in\mathcal{T}_{p}}\texttt{e}^{-K_{2}\left\{ \Gamma_{p,e}-\overrightarrow{n}_{p-1,s(e_{p})}+\overleftarrow{n}_{p,s'(e_{p})}\right\} }.\end{align*}
Let us consider cyclic coordinate decent algorithm which minimizes
$f$ cyclically with respect to the coordinate variable. Thus $\mathbf{m}_{1}$
is changed first, then $\mathbf{m}_{2}$ and so forth through $\mathbf{m}_{N}.$
Then \eqref{eq:JLP3-1}, \eqref{eq:JLP3-2}, and \eqref{eq:JLP3-3}
are equivalent to for\emph{ }each \emph{$p\in\mathcal{I}$ }with proper\emph{
$\mathcal{X}_{p}$ }as

\begin{align*}
 & \min_{\xi_{p}\in\mathcal{X}_{p}}f\left(\mathbf{m}_{1},\,\ldots\,,\,\mathbf{m}_{p-1},\,\xi_{p},\,\mathbf{m}_{p+1},\,\ldots\,,\,\mathbf{m}_{N}\right)\\
= & \min_{\xi_{p}\in\mathcal{X}_{p}}\!\frac{1}{K_{1}}\sum_{j\in\emph{J}}\ln\!\!\sum_{\mathcal{B}\in\mathcal{E}_{j}}\texttt{e}^{\!\!-K_{1}\left\{ \xi_{p,j}\mathbbm{1}_{\mathcal{N}(j)}(p)\mathbbm{1}_{\mathcal{B}}(i)+\!\!\!\sum\limits _{i\in\mathcal{N}(j)}m_{i,j}\mathbbm{1}_{\mathcal{B}}(i)\right\} }\\
+ & \frac{1}{K_{2}}\ln\sum_{e\in\mathcal{T}_{p}}\mbox{\ensuremath{\exp}}\left\{ -K_{2}\left(b_{p,e}-\sum_{j\in\mathcal{N}(p)}\xi_{p,j}\delta_{x(e_{p})=1}\right)\right.\\
+ & \ln\negthinspace\sum_{\left\{ e_{1},\ldots,e_{p-1}\right\} }\negthinspace\negthinspace\texttt{e}^{-K_{2}\left\{ n_{1,s(e_{2})}-\sum\limits _{i=2}^{p-1}b_{i,e}+\sum\limits _{i=2}^{p-1}\sum\limits _{j\in\mathcal{N}(i)}m_{i,j}\delta_{x(e_{i})=1}\right\} }\\
+ & \left.\ln\negthinspace\sum_{\left\{ e_{p+1},\ldots,e_{N}\right\} }\negthinspace\negthinspace\texttt{e}^{-K_{2}\left\{ \sum\limits _{i=p+1}^{N}b_{i,e}-\sum\limits _{i=p+1}^{N}\sum\limits _{j\in\mathcal{N}(i)}m_{i,j}\delta_{x(k_{i})=1}\right\} }\right\} .\end{align*}
Using the properties of log-sum-exp functions (e.g., see \cite[p. 72]{Boyd-2004}),
one can verify that $f$ is continuously differentiable and convex.
The minimum over $\xi_{p}$ for all $i\in\mathcal{I}$ is uniquely
obtained because of the unique KKT solution in Lemma \ref{lem:InnerLoop}.
Therefore,\emph{ }we can apply the Proposition \ref{pro:convergence}
to achieve the desired convergence result under the modified update
schedule. It is worth mentioning that the Hamming weight condition
prevents degeneracy of Problem-DS based on the fact that, otherwise,
some pairs of bits must always be equal. %
\begin{comment}
Since Algorithm \ref{alg:IJLP} can be seen as taking {}``spacer
step'' that does not increase the value of $f$ , we finish by applying
the convergence result for the cyclic method. 
\end{comment}
{}

\subsection{Proof of Lemma \ref{lem:relaxation bound}\label{sub:Proofrelaxationbnd}}

Denote the optimum solution of Problem-P\emph{ }by $\mathbf{g}^{*}$
and $\mathbf{w}^{*}$ and the optimum solution of Problem-PS by $\tilde{\mathbf{g}}$
and $\tilde{\mathbf{w}}.$ Since $\mathbf{g}^{*}$ and $\mathbf{w}^{*}$
are the optimal with respect to the Problem-P\emph{,} we have\emph{
\begin{equation}
P^{*}=\sum_{i\in\mathcal{I}}\sum_{e\in\mathcal{T}_{i}}b_{i,e}g_{i,e}^{*}\leq\sum_{i\in\mathcal{I}}\sum_{e\in\mathcal{T}_{i}}b_{i,e}\tilde{g}_{i,e}=\tilde{P}.\label{eq:leftbound}\end{equation}
}On the other hand, $\tilde{\mathbf{g}}$ and $\tilde{\mathbf{w}}$
are the optimal with respect to the Problem-PS\emph{,} we have\begin{multline*}
\sum_{i\in\mathcal{I}}\sum_{e\in\mathcal{T}_{i}}b_{i,e}\tilde{g}_{i,e}-\frac{1}{K_{1}}\sum_{j\in\mathcal{J}}H(\tilde{w}_{j})-\frac{1}{K_{2}}H(\tilde{g}_{p})\\
\leq\sum_{i\in\mathcal{I}}\sum_{e\in\mathcal{T}_{i}}b_{i,e}g_{i,e}^{*}-\frac{1}{K_{1}}\sum_{j\in\mathcal{J}}H(w_{j}^{*})-\frac{1}{K_{2}}H(g_{p}^{*}),\end{multline*}
where $H(\cdot)$ is the entropy defined by (\ref{eq:entropy_def}).\emph{
}We rewrite this as \begin{align}
 & \sum_{i\in\mathcal{I}}\sum_{e\in\mathcal{T}_{i}}b_{i,e}\tilde{g}_{i,e}\nonumber \\
 & \leq\sum_{i\in\mathcal{I}}\sum_{e\in\mathcal{T}_{i}}b_{i,e}g_{i,e}^{*}+\frac{1}{K_{1}}\left(\sum_{j\in\mathcal{J}}H(\tilde{w}_{j})-\sum_{j\in\mathcal{J}}H(w_{j}^{*})\right)\nonumber \\
 & +\frac{1}{K_{2}}\left(H(\tilde{g}_{p})-H(g_{p}^{*})\right)\nonumber \\
 & \leq\sum_{i\in\mathcal{I}}\sum_{e\in\mathcal{T}_{i}}b_{i,e}g_{i,e}^{*}+\frac{1}{K_{1}}\sum_{j\in\mathcal{J}}H(\tilde{w}_{j})+\frac{1}{K_{2}}H(\tilde{g}_{p}).\label{eq:right1}\end{align}
The last inequality is due to nonnegativity of entropy. Using Jensen's
inequality, we obtain\begin{align}
\sum_{j\in\mathcal{J}}H(\tilde{w}_{j}) & \leq\sum_{j\in\mathcal{J}}\ln\,\left|\mathcal{E}_{j}\right|=\sum_{j\in\mathcal{J}}\left(\left|\mathcal{N}(j)\right|-1\right)\ln\,2\nonumber \\
 & =N\left(1-R+\bar{\mathcal{N}}\right)\ln\,2\label{eq:right2}\end{align}
and\begin{equation}
H(\tilde{g}_{p})\leq\ln\, O.\label{eq:right3}\end{equation}
By substituting \eqref{eq:right2} and \eqref{eq:right3} to \eqref{eq:right1},
we have\begin{equation}
\tilde{P}-P^{*}\leq\frac{N\left(1-R+\bar{\mathcal{N}}\right)\ln\,2}{K_{1}}+\frac{\ln\, O}{K_{2}}.\label{eq:rightbound}\end{equation}
Combining \eqref{eq:leftbound} and \eqref{eq:rightbound} gives the
result. %
{}%
{}

\subsection{Proof of Lemma \ref{lem:Fesability}\label{sub:Prooffeasibility}}

%
\begin{comment}
For\emph{ }$p\in\mathcal{I}$, \foreignlanguage{english}{Problem-D1
is equivalent to}
\end{comment}
{}For the coordinate-descent solution of Problem-DS, minimizing over
the $p$-th block gives \emph{\begin{equation}
-\min_{\left\{ m_{p,j}\right\} _{j\in\mathcal{N}(p)}}\!\!\frac{1}{K_{1}}\sum_{j\in\mathcal{N}(p)}\mathcal{\ln}\sum_{\mathcal{B}\in\mathcal{E}_{j}}\texttt{e}^{-K_{1}\left\{ \sum_{i\in\mathcal{N}(j)}m_{i,j}\mathbbm{1}_{\mathcal{B}}(i)\right\} }\label{eq:JLP3-1-1}\end{equation}
}subject to\emph{\[
\Gamma_{p,e}=\overrightarrow{n}_{p-1,s(e)}-\overleftarrow{n}_{p,s'(e)},\,\forall e\in\mathcal{T}_{p}.\]
}The solution can be obtained by applying the KKT conditions and this
yields\begin{equation}
\frac{\sum_{e:x(e)=1}\lambda_{p,e}}{1-\sum_{e:x(e)=1}\lambda_{p,e}}=\texttt{e}^{K_{1}\left(M_{p,j}-m_{p,j}\right)}.\label{eq:sub1}\end{equation}
Given a feasible solution of the modified Algorithm \ref{alg:IJLP},
we define\begin{eqnarray*}
\lambda_{i}^{j} & \triangleq & \sum_{e:x(e)=1}\lambda_{i,e}^{j}=\frac{1}{1+\texttt{e}^{K_{1}\left(m_{i,j}-M_{i,j}\right)}},\\
\lambda_{i} & \triangleq & \frac{1}{|\mathcal{N}(i)|}\sum_{j\in\mathcal{N}(i)}\lambda_{i}^{j}=\sum_{e:x(e)=1}\lambda_{i,e}\end{eqnarray*}
with\[
\lambda_{i,e}\triangleq\frac{1}{|\mathcal{N}(i)|}\sum_{j\in\mathcal{N}(i)}\lambda_{i,e}^{j}\]
and\[
\epsilon\triangleq\max_{i\in\mathcal{I}}\max_{j\in\mathcal{N}\left(i\right)}|\lambda_{i}^{j}-\lambda_{i}|.\]
Suppose we stop iterating when $\epsilon\leq\frac{1}{6}$ and define\begin{eqnarray*}
\hat{\lambda}_{i} & \triangleq & \left(1-6\epsilon\right)\lambda_{i}+6\epsilon\sum_{e:x(e)=1}\frac{1}{|E|}\\
 & = & \left(1-6\epsilon\right)\lambda_{i}+3\epsilon=\sum_{e:x(e)=1}\hat{\lambda}_{i,e},\end{eqnarray*}
where\[
\hat{\lambda}_{i,e}\triangleq\left(1-6\epsilon\right)\lambda_{i,e}+\frac{6\epsilon}{|E|}.\]
First, we claim that $\mathbf{\hat{\lambda}}\triangleq\left\{ \hat{\lambda}_{i}\right\} \in\mathcal{P}(H).$
This is because setting \begin{equation}
w_{j,\mathcal{B}}\triangleq\frac{\texttt{e}^{-K_{1}\sum_{l\in\mathcal{N}(j)}m_{l,j}\mathbbm{1}_{\mathcal{B}}(l)}}{\sum_{\mathcal{B}'\in\mathcal{E}_{j}}\texttt{e}^{-K_{1}\sum_{l\in\mathcal{N}(j)}m_{l,j}\mathbbm{1}_{\mathcal{B}'}(l)}}\label{eq:w_new}\end{equation}
obviously satisfies for \foreignlanguage{english}{$\forall j\in\mathcal{J}$}
\[
w_{j,\mathcal{B}}\geq0,\,\,\,\forall\mathcal{B}\in\mathcal{E}_{j},\,\,\,\,\sum_{\mathcal{B}\in\mathcal{E}_{j}}w_{j,\mathcal{B}}=1\]
and satisfies for $\forall i\in\mathcal{I},\, j\in\mathcal{N}\left(i\right)$\begin{eqnarray*}
\sum_{\mathcal{B}\in\mathcal{E}_{j},\mathcal{B}\ni i}w_{j,\mathcal{B}}=\!\! & \frac{\sum_{\mathcal{B}\in\mathcal{E}_{j},\mathcal{B}\ni i}\texttt{e}^{-K_{1}\sum_{l\in\mathcal{N}(j)}m_{l,j}\mathbbm{1}_{\mathcal{B}}(l)}}{\sum_{\mathcal{B}'\in\mathcal{E}_{j}}\texttt{e}^{-K_{1}\sum_{l\in\mathcal{N}(j)}m_{l,j}\mathbbm{1}_{\mathcal{B}'}(l)}}=\lambda_{i}^{j}.\end{eqnarray*}
From \cite[p. 4841]{Burshtein-it09}, it follows that $\tilde{\lambda}\in\mathcal{P}(H)$.%
\begin{comment}
Thus by Definition \ref{def:LCP} \[
\sum_{i\in S}\lambda_{i}^{j}-\!\!\!\sum_{i\in L-S}\!\!\lambda_{i}^{j}\leq\left|S\right|-1.\]
For each $S\subseteq L,\,\left|S\right|\text{odd}$, we have\begin{eqnarray*}
\left|S\right|-\sum_{i\in S}\hat{\lambda}_{i}+\!\!\!\sum_{i\in L-S}\!\!\hat{\lambda}_{i} & = & \sum_{i\in S}\left(1-\hat{\lambda}_{i}\right)+\!\!\!\sum_{i\in L-S}\!\!\hat{\lambda}_{i}\\
 & = & 3\epsilon|L|+\left(1-6\epsilon\right)\left[\sum_{i\in L-S}\lambda_{i}+\sum_{i\in S}\left(1-\lambda_{i}\right)\right]\\
 & \geq & 3\epsilon|L|+\left(1-6\epsilon\right)\left[\sum_{i\in L-S}\left(\lambda_{i}^{j}-\epsilon\right)+\sum_{i\in S}\left(1-\lambda_{i}^{j}-\epsilon\right)\right]\\
 & \geq & 3\epsilon|L|+\left(1-6\epsilon\right)\left(1-\epsilon|L|\right)\geq1.\end{eqnarray*}
Again, by Definition \ref{def:LCP}, $\mathbf{\hat{\lambda}}\in\mathcal{P}(H).$ 
\end{comment}
{} Next, we show that $\left\{ \hat{\lambda}_{i,e}\right\} \in\mathcal{T}.$
Note that defining\[
\lambda_{i,e}\triangleq\frac{\texttt{e}^{-K_{2}\left\{ \Gamma_{i,e}-\overrightarrow{n}_{i-1,s(e_{i})}+\overleftarrow{n}_{i,s'(e_{i})}\right\} }}{\sum_{e\in\mathcal{T}_{i}}\texttt{e}^{-K_{2}\left\{ \Gamma_{i,e}-\overrightarrow{n}_{i-1,s(e_{i})}+\overleftarrow{n}_{i,s'(e_{i})}\right\} }}\]
implies that (by \eqref{eq:Update2}) \[
\frac{\sum_{e:x(e)=1}\lambda_{i,e}}{1-\sum_{e:x(e)=1}\lambda_{i,e}}=\texttt{e}^{K_{1}\left(M_{p,j}-m_{p,j}\right)},\]
obviously satisfies \foreignlanguage{english}{for $\forall i\in\mathcal{I}$}
\[
\lambda_{i,e}\geq0,\,\,\,\forall e\in\mathcal{T}_{i},\,\,\,\sum_{\substack{e\in\mathcal{T}_{i}}
}\lambda_{i,e}=1\]
and for $\forall i\in\mathcal{I}\setminus N,\,\,\, k\in\emph{S}$
by \eqref{eq:JLP3-2} and \eqref{eq:JLP3-3}\begin{eqnarray*}
\sum_{e:s'(e)=k}\lambda_{i,e} & = & \frac{\sum_{e:s'(e)=k}\texttt{e}^{-K_{2}\left\{ \Gamma_{i,e}-\overrightarrow{n}_{i-1,s(e_{i})}+\overleftarrow{n}_{i,s'(e_{i})}\right\} }}{\sum_{e\in\mathcal{T}_{i}}\texttt{e}^{-K_{2}\left\{ \Gamma_{i,e}-\overrightarrow{n}_{i-1,s(e_{i})}+\overleftarrow{n}_{i,s'(e_{i})}\right\} }}\\
 & = & \sum_{e:s(e)=k}\lambda_{i+1,e}.\end{eqnarray*}
Furthermore, \begin{eqnarray*}
\sum_{\substack{e\in\mathcal{T}_{i}}
}\hat{\lambda}_{i,e} & = & \left(1-6\epsilon\right)\sum_{\substack{e\in\mathcal{T}_{i}}
}\lambda_{i,e}+6\epsilon\sum_{\substack{e\in\mathcal{T}_{i}}
}\frac{1}{|E|}=1,\end{eqnarray*}
\begin{eqnarray*}
\sum_{\substack{e:s'(e)=k}
}\hat{\lambda}_{i,e} & = & \left(1-6\epsilon\right)\sum_{\substack{e:s'(e)=k}
}\lambda_{i,e}+6\epsilon\sum_{\substack{e:s'(e)=k}
}\frac{1}{|E|}\\
 & = & \left(1-6\epsilon\right)\sum_{\substack{e:s(e)=k}
}\lambda_{i+1,e}+6\epsilon\sum_{\substack{e:s(e)=k}
}\frac{1}{|E|}\\
 & = & \sum_{\substack{e:s(e)=k}
}\hat{\lambda}_{i+1,e},\end{eqnarray*}
and by Definition \ref{def:trellis-polytope}, $\mathbf{\hat{\lambda}}\in\mathcal{P}(H)$.
\foreignlanguage{english}{Therefore, we conclude that $\left\{ \hat{\lambda}_{i,e}\right\} \in\mathcal{P}_{\mathcal{T}}(H)$
is feasible in Problem-P.} From \cite[p. 4855]{Burshtein-it09}, it
follows that there exist feasible $\hat{w}_{j}$ vectors associated
with $\left\{ \hat{\lambda}_{i,e}\right\} $. %
{}

Denote the minimum value of Problem-PS by $\tilde{P}.$ Then by the
Lagrange duality we can upper bound $\tilde{P}_{\epsilon}-\tilde{P}$
with \begin{align*}
\sum_{i\in\mathcal{I}} & \sum_{e\in\mathcal{T}_{i}}b_{i,e}\hat{\lambda}_{i,e}-\frac{1}{K_{1}}\sum_{j\in\mathcal{J}}H(\hat{w}_{j})-\frac{1}{K_{2}}H(\hat{\lambda}_{p})-\tilde{P}\\
\leq & \sum_{i\in\mathcal{I}}\sum_{e\in\mathcal{T}_{i}}b_{i,e}\hat{\lambda}_{i,e}-\frac{1}{K_{1}}\sum_{j\in\mathcal{J}}H(\hat{w}_{j})-\frac{1}{K_{2}}H(\hat{\lambda}_{p})\\
 & +\frac{1}{K_{1}}\sum_{j\in\mathcal{J}}\mathcal{\ln}\sum_{\mathcal{B}\in\mathcal{E}_{j}}\texttt{e}^{-K_{1}\left\{ \sum_{i\in\mathcal{N}(j)}m_{i,j}\mathbbm{1}_{\mathcal{B}}(i)\right\} }\\
\stackrel{(a)}{\leq} & \frac{1}{K_{1}}\!\sum_{j\in\mathcal{J}}\!\left[H(w_{j})\!-\! H(\hat{w}_{j})\right]\!-\!\frac{1}{K_{2}}\! H(\hat{\lambda}_{p})\\
 & +\!\epsilon\!\left(\!3\!\sum_{l\in\mathcal{I}}\!\sum_{\substack{e\in\mathcal{T}_{l}}
}\!\left|b_{l,e}\right|\!+\! CN\!\right)\\
\leq & \frac{1}{K_{1}}\sum_{j\in\mathcal{J}}H(w_{j})+\epsilon N\left(\!\frac{3}{N}\!\sum_{l\in\mathcal{I}}\!\sum_{\substack{e\in\mathcal{T}_{l}}
}\left|b_{l,e}\right|+C\right),\end{align*}
where $(a)$ is given by rewriting \eqref{eq:w_new} as\begin{align*}
 & \frac{1}{K_{1}}\sum_{j\in\mathcal{J}}\mathcal{\ln}\sum_{\mathcal{B}\in\mathcal{E}_{j}}\texttt{e}^{-K_{1}\left\{ \sum_{i\in\mathcal{N}(j)}m_{i,j}\mathbbm{1}_{\mathcal{B}}(i)\right\} }\\
 & =\frac{1}{K_{1}}\sum_{j\in\mathcal{J}}H(w_{j})-\sum_{j\in\mathcal{J}}\sum_{\mathcal{B}\in\mathcal{E}_{j}}w_{j,\mathcal{B}}\sum_{l\in\mathcal{N}(j)}m_{l,j}\mathbbm{1}_{\mathcal{B}}(l)\\
 & \leq\frac{1}{K_{1}}\!\sum_{j\in\mathcal{J}}\! H(w_{j})\!-\!\sum_{l\in\mathcal{I}}\!\sum_{\substack{e\in\mathcal{T}_{l}}
}\! b_{l,e}\hat{\lambda}_{l,e}\!+\!\epsilon\!\left(\!3\!\sum_{l\in\mathcal{I}}\!\sum_{\substack{e\in\mathcal{T}_{l}}
}\!\left|b_{l,e}\right|\!+\! CN\!\right)\!.\end{align*}
The last step of this equation follows from\begin{align*}
\sum_{j\in\mathcal{J}} & \sum_{\mathcal{B}\in\mathcal{E}_{j}}w_{j,\mathcal{B}}\sum_{l\in\mathcal{N}(j)}m_{l,j}\mathbbm{1}_{\mathcal{B}}(l)\\
 & =\sum_{l\in\mathcal{I}}\sum_{j\in\mathcal{N}(l)}m_{l,j}\lambda_{l}^{j}\\
 & \geq\sum_{l\in\mathcal{I}}\sum_{j\in\mathcal{N}(l)}m_{l,j}\left(\lambda_{l}-\epsilon\right)\\
 & \geq\sum_{l\in\mathcal{I}}\!\sum_{\substack{e\in\mathcal{T}_{l}}
}\left(\delta_{x(e)=1}\!\sum_{j\in\mathcal{N}(l)}m_{l,j}\right)\!\lambda_{l,e}\!-\!\epsilon\sum_{l\in\mathcal{I}}\!\sum_{j\in\mathcal{N}(l)}\left|m_{l,j}\right|\\
 & \geq\sum_{l\in\mathcal{I}}\sum_{\substack{e\in\mathcal{T}_{l}}
}b_{l,e}\lambda_{l,e}-\epsilon CN\\
 & \geq\sum_{l\in\mathcal{I}}\sum_{\substack{e\in\mathcal{T}_{l}}
}b_{l,e}\hat{\lambda}_{l,e}-3\epsilon\sum_{l\in\mathcal{I}}\sum_{\substack{e\in\mathcal{T}_{l}}
}\left|b_{l,e}\right|-\epsilon CN.\end{align*}
In the above equation, the details of the last two inequalities are
not included due to space limitations, but they can be derived using
arguments very similar to \cite[p. 4840-4841]{Burshtein-it09}.

%
\begin{comment}
For any $\delta>0$\emph{,} after sufficiently many iterations with
sufficiently large $K_{1}$ and $K_{2}$\emph{ }until $\epsilon$
becomes sufficiently small, the modified Algorithm \ref{alg:IJLP}
outputs $\mathbf{\tilde{g}_{\epsilon}}$ with the minimum value $\tilde{P}_{\epsilon}$
which satisfies \[
\frac{\tilde{P}_{\epsilon}-\tilde{P}}{N}\leq\frac{\delta}{2}\]
by Lemma \ref{lem:convergence}. By combining with Lemma \ref{lem:relaxation bound},
the final $\mathbf{\tilde{g}_{\epsilon}}$ satisfies the given relaxation
bound as \[
\frac{\tilde{P}_{\epsilon}-P^{*}}{N}=\frac{\tilde{P}_{\epsilon}-\tilde{P}}{N}+\frac{\tilde{P}-P^{*}}{N}\leq\delta.\]

\end{comment}
{}

%\bibliographystyle{IEEEtran}
%\addcontentsline{toc}{section}{\refname}\bibliography{IEEEabrv,WCLabrv,WCLbib,WCLnewbib}

\end{document}